\pdfoutput=1
\documentclass[fleqn,usenatbib]{mnras}

\usepackage{graphicx}	
\usepackage{amsmath}	
\usepackage{amssymb}	
\usepackage{mathtools}
\usepackage[utf8]{inputenc}
\usepackage[english]{babel}
\usepackage[section]{placeins}
\usepackage{textcomp,gensymb}
\usepackage{multirow}
\usepackage{hyperref}
\usepackage{xspace}
\usepackage{float}
\usepackage{verbatim}
\usepackage{subcaption}
\usepackage{CJKutf8}

\newcommand{\specialcell}[2][c]{%
  \begin{tabular}[#1]{@{}c@{}}#2\end{tabular}}

\newcommand{\reffig}[1]{Fig.~\ref{#1}}
\newcommand{\refsec}[1]{\S~\ref{#1}}

\newcommand{\orcid}[2]{\href{http://orcid.org/#2}{#1}}

\newcommand{\re}{$\mathrm{R_e}$\xspace}

\defcitealias{wu+2018}{W18}

\def\fe4383{$\mathrm{Fe4383}$\xspace}
\def\Dfe4383{$\Delta \mathrm{Fe4383}$\xspace}
\def\hdeltaa{$\mathrm{H\delta_A}$\xspace}

\def\hgammaa{$\mathrm{H\gamma_A}$\xspace}
\def\dn4000{$\mathrm{D_n}4000$\xspace}

\title[Inverse age gradients of PSB galaxies]{Inverse stellar population age gradients of post-starburst galaxies at $z = 0.8$ with LEGA-C}

\author[\href{http://orcid.org/0000-0003-2388-8172}{F.~D'Eugenio}~et al.]{\parbox{\textwidth}{
\orcid{Francesco D'Eugenio}{0000-0003-2388-8172}$^{1}$\thanks{E-mail: francesco.deugenio@gmail.com},
\orcid{Arjen van der Wel}{0000-0002-5027-0135}$^{1,2}$,
\orcid{Po-Feng Wu~\begin{CJK*}{UTF8}{bkai}(吳柏鋒)\end{CJK*}}{0000-0002-9665-0440}$^{3,2}$,
\orcid{Tania M. Barone}{0000-0002-2784-564X}$^{4,5,6}$,
\orcid{Josha van Houdt}{0000-0003-1888-3705}$^{2}$,
\orcid{Rachel Bezanson}{0000-0001-5063-8254}$^{7}$,
\orcid{Caroline M. S. Straatman}{0000-0001-5937-4590}$^{1}$,
\orcid{Camilla Pacifici}{0000-0003-4196-0617}$^{8}$,
\orcid{Adam Muzzin}{0000-0002-9330-9108}$^{9}$,
\orcid{Anna Gallazzi}{0000-0002-9656-1800}$^{10}$, 
\orcid{Vivienne Wild}{0000-0002-8956-7024}$^{11}$,
\orcid{David Sobral}{0000-0001-8823-4845}$^{12}$,
\orcid{Eric F. Bell}{0000-0002-5564-9873}$^{13}$,
\orcid{Stefano Zibetti}{0000-0003-1734-8356}$^{10}$
\orcid{Lamiya Mowla}{0000-0002-8530-9765}$^{14}$
and
\orcid{Marijn Franx}{0000-0002-8871-3026}$^{15}$}
\vspace{0.4cm}
\\
\parbox{\textwidth}{
$^{1}$Sterrenkundig Observatorium, Universiteit Gent, Krijgslaan 281 S9, B-9000 Gent, Belgium\\
$^{2}$Max-Planck-Institut f\"ur Astronomie, K\"onigstuhl 17, D-69117, Heidelberg, Germany\\
$^{3}$National Astronomical Observatory of Japan, 2-21-1 Osawa, Mitaka, Tokyo 181-8588, Japan\\
$^{4}$Research School of Astronomy and Astrophysics, Australian National University, Canberra, ACT 2611, Australia\\
$^{5}$Sydney Institute for Astronomy, School of Physics, The University of Sydney, NSW, 2006, Australia\\
$^{6}$ARC Centre of Excellence for All Sky Astrophysics in 3 Dimensions (ASTRO 3D), Australia\\
$^{7}$Department of Physics and Astronomy and PITT PACC, University of Pittsburgh, Pittsburgh, PA 15260, USA\\
$^{8}$Space Telescope Science Institute, 3700 San Martin Drive, Baltimore, MD 21218, USA\\
$^{9}$Department of Physics and Astronomy, York University, 4700 Keele St., Toronto, Ontario, M3J 1P3, Canada\\
$^{10}$INAF-Osservatorio Astrofisico di Arcetri, Largo Enrico Fermi 5, I-50125 Firenze, Italy\\
$^{11}$School of Physics and Astronomy, University of St Andrews, North Haugh, St Andrews KY16 9SS, UK (SUPA)\\
$^{12}$Department of Physics, Lancaster University, Lancaster LA1 4YB, UK\\
$^{13}$Department of Astronomy, University of Michigan, Ann Arbor, MI 48109, USA\\
$^{14}$Astronomy Department, Yale University, New Haven, CT 06511, US\\
$^{15}$Leiden Observatory, P.O. Box 9513, 2300 RA, Leiden, The Netherlands\\
}
}

\date{Accepted 2020 Jun 29. Received 2020 June 23; in original form 2020 March 25}

\pubyear{2020}

\begin{document}
\label{firstpage}
\pagerange{\pageref{firstpage}--\pageref{lastpage}}
\maketitle

\begin{abstract}

  We use deep, spatially resolved spectroscopy from the LEGA-C Survey to study
  radial variations in the stellar population of 17 spectroscopically-selected
  post-starburst (PSB) galaxies.
  We use spectral fitting to measure two Lick indices, \hdeltaa and \fe4383, and
  find that, on average, PSB galaxies have radially decreasing \hdeltaa and
  increasing \fe4383 profiles. In contrast, a control sample of quiescent, non-PSB
  galaxies in the same mass range shows outwardly increasing \hdeltaa and decreasing
  \fe4383. The observed gradients are weak ($\approx -0.2 \text{\AA}/R_e$), mainly
  due to seeing convolution. A two-SSP model suggests intrinsic gradients are as
  strong as observed in local PSB galaxies ($\approx -0.8 \text{\AA}/R_e$).
  We interpret these results in terms of inside-out growth (for the
  bulk of the quiescent population) vs star formation occurring last in the centre
  (for PSB galaxies). At $z \approx 0.8$, central starbursts are often the
  result of gas-rich mergers, as evidenced by the high fraction of PSB galaxies
  with disturbed morphologies and tidal features (40\%).
  Our results provide additional evidence for multiple paths to quiescence: a
  standard path, associated with inside-out disc formation and with gradually decreasing
  star-formation activity, without fundamental structural transformation, and a
  fast path, associated with centrally-concentrated starbursts, leaving an inverse
  age gradient and smaller half-light radius.

\end{abstract}




\section{Introduction}\label{s.i}

At any given time, star-forming (SF) galaxies form a sequence in the mass-size
plane \citetext{\citealp{shen+2003}, \mbox{\citealp{vanderwel+2014}}}; this mass-size relation is such that,
at fixed stellar mass, SF galaxies are systematically larger than
non-star-forming galaxies (hereafter: quiescent, or Q galaxies). Moreover, the
average size of both SF and Q galaxies increases with cosmic time \citep{
vanderwel+2009, fagioli+2016, williams+2017}.
Given these properties, it is reasonable to assume that galaxies that have
recently become quiescent have approximately the same size as SF galaxies of the
same mass. This expectation is indeed consistent with the finding that, at fixed
stellar mass, the youngest Q galaxies are also the largest \citep{wu+2018}.

There is however a class of objects, called post-starburst (PSB) galaxies, that
have recently become quiescent, yet contrary to the above expectations are both:
(i) smaller than coeval Q galaxies, and (ii) much smaller than coeval SF
galaxies \citetext{e.g. \citealp{whitaker+2012}, \citealp{almaini+2017},
\citealp{wu+2018}}.
Observationally, PSB galaxies \citep{dressler+gunn1983, couch+sharples1987}
present strong Balmer-line absorption (typical of young, $0.3 - 1 \,
\mathrm{Gyr}$-old stars) but lack $\rm{H\alpha}$ emission (which excludes
recent, $\lesssim 10 \, \mathrm{Myr}$ star formation). Together, these two
properties suggest that PSB galaxies stopped forming stars both \emph{rapidly}
(faster than $\approx 1 \, \mathrm{Gyr}$) and \emph{recently} (within $\approx
1 \, \mathrm{Gyr}$ of their look-back time). The empirical conjunction of
compact structure with a rapid and recent transition to quiescence, suggests
that PSB galaxies followed a
special evolutionary path, either an extreme version of normal galaxy evolution,
or some entirely different channel.

In the local Universe, PSB galaxies are empirically associated either with
galaxy mergers \citep[e.g.][]{zabludoff+1996, bekki+2001, yang+2004, goto2005,
yang+2008, pracy+2009, wild+2009, pawlik+2018} or with ram-pressure stripping in
dense environments \citep{dressler+1999, poggianti+1999, tran+2004,
poggianti+2009, paccagnella+2019}. They show a range of kinematic properties:
from dispersion-dominated kinematics reminiscent of quiescent galaxies
\citep[and consistent with the outcome of mergers;][]{hiner+canalizo2015} to
rotation-supported systems \citep[e.g.][]{norton+2001, pracy+2013, owers+2019}.
\citet{chen+2019} find that stellar kinematics depend on the location of the PSB
regions within the target galaxy.
In any case, even accounting for their relatively short visibility time, local
PSB galaxies represent a marginal mode of galaxy evolution \citep{rowlands+2018}.

However, in the high-redshift Universe, PSB galaxies could be different.
Firstly, it appears that the fraction of PSB galaxies increases with cosmic time
\citetext{e.g. \citealp{dressler+1999}, \citealp{poggianti+1999},
\citealp{wild+2016}, but see e.g. \citealp{balogh+1997}, \citealp{balogh+1999},
\citealp{muzzin+2012} for a different view}. Even if the fraction of PSB
galaxies stayed constant, dense environments become rarer with increasing
redshift \citep[e.g.][]{carlberg+1997, younger+2005}, hence the physics
underlying low- and high-redshift PSB galaxies could be different. This is not
a surprising possibility, because the definition of PSB galaxies is
purely empirical, and different quenching mechanisms could ostensibly leave
similar or identical signatures. In addition, there is evidence for different structural
properties between $z<1$ and $z>1$ PSB galaxies \citep{maltby+2018}, suggesting
that redshift evolution might involve different physical processes.

A possible explanation of the observed properties of PSB galaxies is a
central starburst in a previously normal galaxy. A significant amount of
star formation inside $\approx 1 \, \mathrm{kpc}$ from the centre of a galaxy can
reduce its previous half-light radius, thus explaining the small observed size of
PSB galaxies \citep{wu+2020}. At the same time, central starbursts are
likely to undergo rapid quenching: either because of strong feedback, or because
of the short dynamical time in the central regions of galaxies, which leads
to rapid consumption of the cold gas reservoir \citep[e.g.][]{wang+2019}.
However, without knowledge of the progenitors of PSB galaxies, it is impossible
to establish whether they have always been compact, or if their half-light radii
have become smaller as a result of a central starburst and subsequent quenching.

Still, if the second hypothesis is true, we expect high-z PSB galaxies to exhibit
clear evidence of a central starburst, such as outwardly-increasing stellar age
\citep[as indeed observed in some local PSB galaxies; see e.g.][]{pracy+2013, owers+2019,
chen+2019}. These inverse gradients are contrary to what is observed in the
majority of both SF and Q galaxies: there is in fact overwhelming evidence that most
galaxies form in an inside-out fashion. Firstly, by comparing the size of the
star-forming gas disc to the size of the stellar disc, the \emph{instantaneous}
radial growth rate of SF galaxies has been shown to be positive \citep
{pezzulli+2015, wang+2019, nelson+2016, paulino-afonso+2017, suzuki+2019}.
Secondly, the stellar populations of most SF and Q galaxies have negative age
gradients with radius \citep[e.g.][]{gonzalezdelgado+2015, zibetti+2017}: these
gradients are qualitatively consistent with the outcome of inside-out growth
\emph{integrated} over cosmic time \citep{schonrich+mcmillan2017}\footnote{
Contrary to this picture, there is considerable evidence that some Q galaxies
have positive or U-shaped age gradients \citep{labarbera+2012, zibetti+2020}. As
it will become clear, this fact by itself does not contradict the picture we
present in this paper.}.
This is also true for individually-measured stars in the Milky Way, both overall
\citep[i.e. the bulge is older than the disc, e.g.][]{valenti+2013} and within
the disc itself \citep{martig+2016}.

Measuring age gradients requires high-quality, spatially resolved spectroscopy
in the optical rest-frame, but until now these observations at intermediate/high
redshift have been out of reach, or limited to small samples \citep{belli+2017}.

The Large Early Galaxy Astrophysics Census \citep[LEGA-C;][]{vanderwel+2016}
changed this state of affairs: LEGA-C provides the Astrophysics community with
deep spectra for $\gtrsim 3000$ galaxies at redshift $z\approx0.8$, when the
Universe was only half its present age.
In this work, we leverage the extraordinary depth of LEGA-C to study the
structural imprint of inside-out or central-starburst growth in PSB galaxies at
roughly half the age of the Universe. After introducing the data
and the sample (\refsec{s.da}), we show that PSB galaxies have distinctive gradients
in their Lick indices, different from the control sample (\refsec{s.r}), and
consistent with an inverse age gradient (\refsec{s.2ssp}). We conclude this
work with a discussion of the
implications and with a summary of our findings (\refsec{s.ds}).
Throughout this paper, we assume a flat $\Lambda$CDM Cosmology with $H_0 = 70 \;
\mathrm{km \, s^{-1} \, Mpc^{-1}}$ and $\Omega_m = 0.3$ and a Chabrier
initial-mass function \mbox{\citep{chabrier2003}}. All magnitudes are in the AB
system \citep{oke+gunn1983}.

\section{Data Analysis}\label{s.da}

\subsection{The LEGA-C Survey}\label{s.da.ss.legac}

LEGA-C \citep{vanderwel+2016} is a deep spectroscopic
survey targeting $0.6 < z < 1.0$ massive galaxies in the COSMOS field, using the
VIMOS spectrograph \citep{lefevre+2003} on the ESO Very Large Telescope.
The LEGA-C primary sample consists of $\approx 3000$ galaxies brighter than $K_s = 20.7
- 7.5 \log[(1+z)/1.8]$, roughly equivalent to a mass-selection limit $\log M_\star /
\mathrm{M_\odot} > 10$. Each galaxy was observed for $20 \, \mathrm{h}$,
reaching a typical signal-to-noise ratio $SNR\approx20\,\text{\AA}^{-1}$ in the
continuum. The median seeing full-width at half-maximum is $\mathrm{FWHM} = 1.0
\, \mathrm{arcsec}$, sampled with $0.205 \, \mathrm{arcsec}$ spatial pixels,
which is sufficient to extract spatial information
\citetext{e.g. \citealp{bezanson+2018}; hereafter we refer to spatial pixels as
spaxels and to spectral pixels simply as pixels}.
What sets LEGA-C apart from previous surveys is its unique combination of ultra-deep
data, large sample size and spatial resolution: these three characteristics
enable us to study resolved stellar population properties in a statistically
meaningful sample.

We use stellar masses $M_\star$ measured by fitting the 30 photometric bands of
the UltraVISTA catalogue, covering the wavelength range $0.15-24 \, \mathrm{\mu m}$
\citep{muzzin+2013a}. Semi-major axis effective radii $R_e$ were measured on
\textit{HST} ACS F814W images obtained as part of the COSMOS program \citep{scoville+2007b},
using {\sc galfit} \citep{peng+2010a} and the procedure of \citet{vanderwel+2012}.
Spectroscopic redshifts were obtained by fitting the galaxy spectra
with a library of synthetic stellar population models (Conroy et al., in prep.),
following the procedure highlighted in \citet{bezanson+2018}.
Rest-frame $U-V$ and $V-J$ colours were calculated by fitting a set of seven
template spectra to the UltraVISTA SED photometry, \citep[see][for a full
description]{straatman+2018}.

\subsection{Sample selection}\label{s.da.ss.samp.sel}

Our parent sample is taken from the LEGA-C public Data Release 2
\citep{straatman+2018}, selected to have $f_{use}=1$ \citetext{1462 galaxies; see
\citealp{straatman+2018} for the definition of the quality flag $f_{use}$}, with four or more
radial bins with $SNR > 10 \, \mathrm{pixel}^{-1}$ each (see \S\ref{s.da.ss.lick}; 614 galaxies),
and with $R_e>0.5 \times (\mathrm{FWHM}$/2.355), where the seeing
FWHM was measured directly on the slit images of each galaxy, using \textit{HST}
photometry as unconvolved reference (van~Houdt~in~prep.; 603 galaxies).
We identify passive galaxies in this sample using the $U-V$ vs $V-J$
colour-colour diagram \citetext{cf. \citealp{labbe+2005}, \citealp{vanderwel+2016};
298 galaxies}, and
we select 17 PSB galaxies\footnote{The total was 19 galaxies, but we further
discarded two targets contaminated by interlopers.} as having a median index
over the spatial measurements $\mathrm{H\delta_A} \geq 4\, \text{\AA}$,
corresponding to an approximate simple-stellar-population-age of $1-1.5
\, \mathrm{Gyr}$, and typically adopted as selection threshold for
spectroscopically-selected PSB galaxies \citep{wu+2018}\footnote{The selection
is almost unchanged (one galaxy is changed) if we use the value of the
integrated \hdeltaa, as expected from the excellent agreement between the median
and integrated index measurements (see Fig.~\ref{f.da.index.comp}). Moreover, we
note that our results are qualitatively unchanged if we use a stricter selection
threshold at $\mathrm{H\delta_A} \geq 5 \, \text{\AA}$.}
The combination of colour and absorption-strength
selection criteria is robust against contamination from dust-obscured starbursts
\citetext{see e.g. \citealp{wu+2018}, and \citealp{dressler+1999, poggianti+1999}}.
Other authors have used a cut on inclination \citep{pawlik+2018}, but we find that
inclination does not drive our results, hence no inclination cut has been applied
(see \S\ref{s.r.ss.caveats}).
These PSB galaxies consist of six centrals, six isolated, three satellites and
two where no environment could be assigned \citep{darvish+2017}. As such, we can
exclude that this sample is dominated by satellites, or subject to ram-pressure
stripping.
As a control sample, we take 141 passive galaxies having the same mass \emph{range}
as the PSB sample, but median $\mathrm{H\delta_A} < 4 \, \text{\AA}$. Choosing
a stricter cut in \hdeltaa does not change the properties of the control sample,
because the bulk of the control galaxies have \hdeltaa well below $4 \,
\text{\AA}$ (only 14 galaxies have median \hdeltaa between $2.5$ and $4.0 \,
\text{\AA}$). Even though a control sample having the same mass
\emph{distribution} as the PSB sample would be better suited to control for
mass-related biases, in practice such selection is not possible with our data
(see \S\ref{s.r.ss.caveats} for a discussion).

\begin{figure}
  \includegraphics[type=pdf,ext=.pdf,read=.pdf,width=1.0\columnwidth]{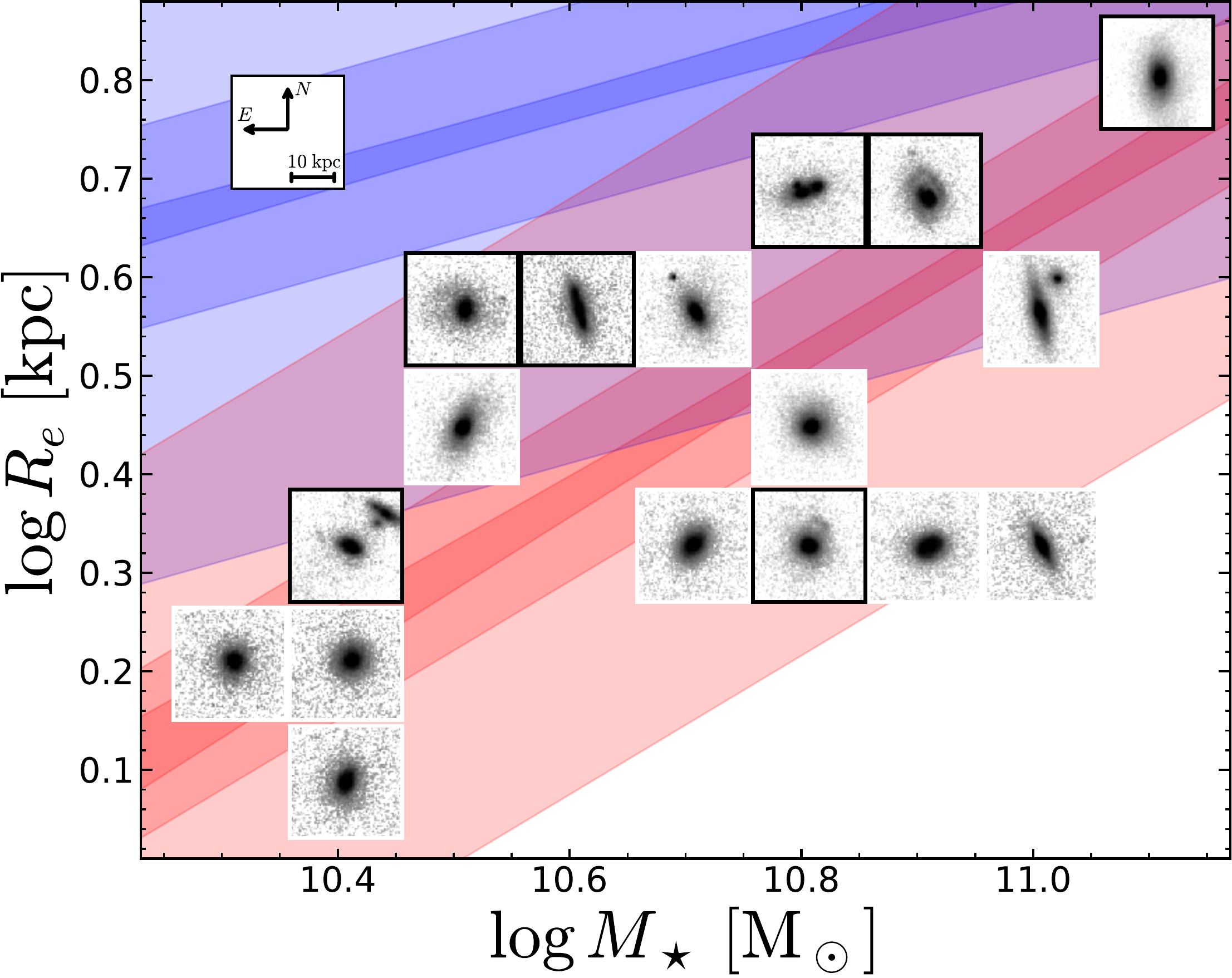}
  \caption{Our sample of post-starburst galaxies has a range of masses, sizes
  and morphologies, including $40\%$ of mergers. Galaxies with 
  $P(\mathrm{merger}) \geq 0.5$ are highlighted using solid black contour insets.
  Each image is a $10\times10 \, \mathrm{arcsec^2}$ cutout from \textit{HST}
  ACS F814W, and the inset centre is placed at the approximate location on the
  mass-size plane of the portrayed galaxy (offsets of up to $0.2 \, \mathrm{dex}$
  are allowed for display purposes).
  The orientation and median physical scale of the images is indicated in the
  top left inset. The red (blue) transparent regions indicate the best-fit linear
  model to the mass-size relation for quiescent (star-forming) galaxies; from
  darkest to lightest, the regions highlight the 95\% confidence interval, the
  68\% prediction interval and the 95\% prediction interval. The best-fit
  parameters for the mass-size relations were derived using the least-trimmed
  squares algorithm \citep{rousseeuw+driessen2006, cappellari+2013a}. Despite
  our spatial
  resolution constraint, which systematically selects the largest PSB galaxies,
  the sample lies appreciably below the star-forming mass-size relation for LEGA-C
  }\label{f.da.hstimg}
\end{figure}

The position of the PSB galaxies on the mass-size plane is illustrated in
\reffig{f.da.hstimg}, where each target is represented by its \textit{HST}
image, so that each inset is placed at the approximate location of the galaxy
portrayed (each inset was allowed a maximum offset of $0.2 \, \mathrm{dex}$ in
$\log M_\star$ and $0.1 \, \mathrm{dex}$ in $\log R_e$, to avoid overlappings).

We use \textit{HST} imaging to assign to each galaxy a probability that it
underwent a recent merger. Three astronomers visually inspected the galaxies and
the residuals of the best-fit {\sc galfit} models, looking for two merger
signatures: tidal features and double cores. Notice that close neighbours are
not classified as mergers, unless tidal features are visible either in the
\textit{HST} image or in the residuals. Galaxies were classified as either
mergers (score of 1) or non-merger (score of 0). The average score is the
probability that a given target is a merger remnant. Galaxies with a score
$P(\mathrm{merger}) \geq 0.5$
are highlighted by insets with solid black contours in \reffig{f.da.hstimg}. For
PSB galaxies, we find 7/17 or 40\% of mergers.

Galaxies with/without detectable merger signatures have consistent values of
the integrated \hdeltaa (mean $\langle \mathrm{H\delta_A} \rangle = 5.91\pm0.45\,\AA$
and $5.49 \pm 0.28 \, \AA$ respectively) and \fe4383 (mean $\langle
\mathrm{Fe4383} \rangle = 1.75 \pm 0.36 \, \AA$ and $2.34 \pm 0.37 \, \AA$
respectively). However, we find that galaxies with integrated
$\mathrm{H\delta_A}$ larger than the median value ($\mathrm{H\delta_A} \geq
5.54 \, \AA$), have somewhat smaller size than galaxies with $\mathrm{H\delta_A}
< 5.54 \, \AA$), but the significance is only two standard deviations. Still,
the direction of this anti-correlation between \hdeltaa and half-light radius
is the same reported in \citet{wu+2020} for a larger PSB sample.
Nevertheless, we find that splitting the sample at $P(\mathrm{merger})\geq0.5$
or at the median value of the half-light radius does not change our results,
apart from lowering their statistical significance (\S\ref{s.r.ss.caveats}).

\subsection{Spatially resolved Lick index measurements}\label{s.da.ss.lick}

We measure Lick indices, defined as in \citet{worthey+ottaviani1997} and
\citet{trager+1998}, as well as the \dn4000 index \citep{balogh+1999}.
The method we use, developed by \citet{scott+2017} and \citet{barone+2020}, can be
thought of as a non-parametric emission-line subtraction.
The goal of their algorithm is to leverage spectral
information away from emission-line regions to  reconstruct the galaxy stellar spectrum
inside such regions. Empirical stellar spectra have been shown to encode significantly more
information per spectral element compared to synthetic spectra \citep[e.g.][]{
martins+coelho2007, plez2011}, therefore
they are more likely to accurately reproduce the observed galaxy spectra
\citep[e.g.][their fig.~25]{vandesande+2017}, and to capture the necessary information to
reconstruct the spectrum in the masked regions.
For this reason, we fit the LEGA-C spectra with an empirical stellar template library.
We use the MILES stellar template library because of its generous range in stellar
classes \citep{sanchez-blazquez+2006, falcon-barroso+2011}, but we obtain equivalent
results using the high spectral resolution ELODIE library \citep{prugniel+soubiran2001}.\\

\begin{figure*}
\begin{minipage}{1.0\textwidth}
  \includegraphics[type=pdf,ext=.pdf,read=.pdf,width=1.0\textwidth]{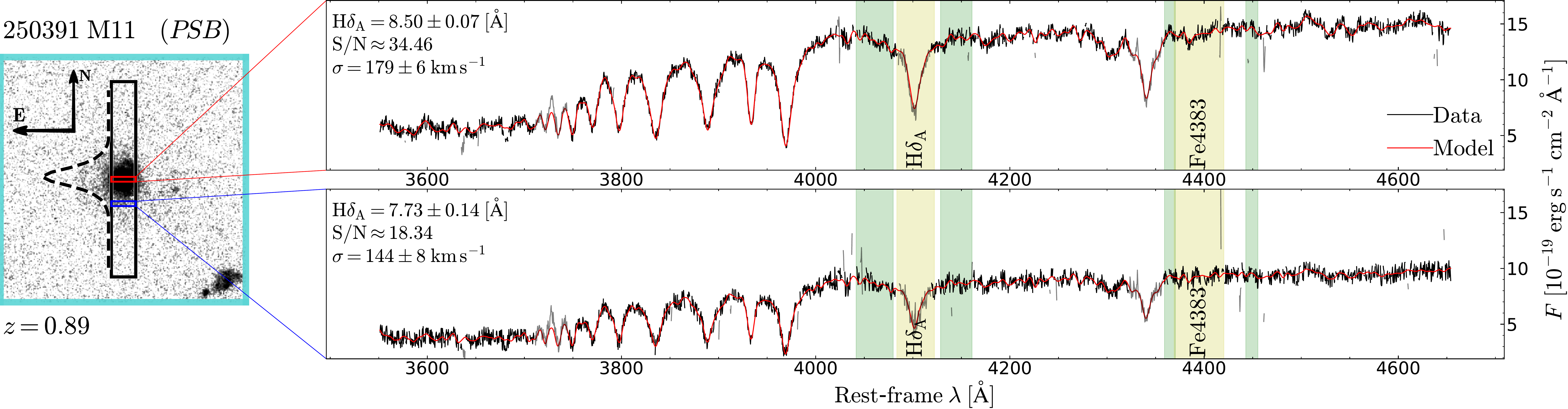}
    \end{minipage}

    \vspace*{0cm} 

    \begin{minipage}{1.0\textwidth}
  \includegraphics[type=pdf,ext=.pdf,read=.pdf,width=1.0\textwidth]{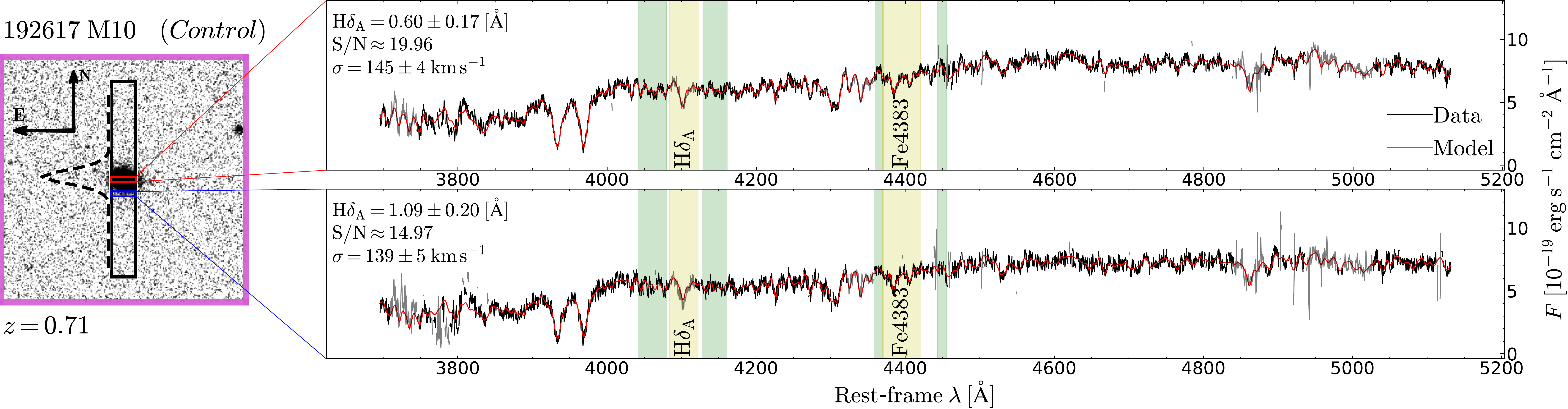}
    \end{minipage}
  \caption{Central and outer rest-frame spectra for the post-starburst galaxy
  M11.250391 (top) and for the quiescent galaxy M10.192617 (bottom).
  The continuum and the index regions for the Lick indices \hdeltaa and \fe4383
  are highlighted in
  green and yellow respectively. The approximate spaxel position for each spectrum
  is marked on the finding chart. For the quiescent galaxy, \hdeltaa is higher
  in the outskirts than in the centre, for the PSB galaxy \hdeltaa is strongest
  in the centre. The black lines represent LEGA-C data, whereas the red lines are
  the {\sc pPXF} best-fit models (see \S\ref{s.da.ss.lick} for more details).
  Regions of the spectrum where the data were masked are rendered
  in grey: these regions are excluded either because they fail a three $\sigma$
  clipping threshold, or because of possible emission lines (regardless of
  whether line emission has been detected).
  }\label{f.da.fitsnrall}
\end{figure*}

We fit the stellar continuum using the penalised Pixel Fitting code {\sc pPXF} 
\citep{cappellari+emsellem2004}, following the procedure developed by \citet{scott+2017}. 
In short, we fit the spectrum optimising for the template weights, for the first and
second moment of the line-of-sight velocity distribution, $v$ and $\sigma$, and for a 
12\textsuperscript{th}-order additive polynomial.\footnote{Even though this choice
is motivated by convergence criteria (i.e. higher degree polynomials do not exhibit
faster variation with wavelength), we find that the value of the Lick indices does
not change for degrees higher than 2. This behaviour is due to a combination of
the local nature of the Lick measurements and of our good data quality. Replacing
additive polynomials with multiplicative polynomials does not change our results
either \citep[see also][]{bezanson+2018}.}

The fit is performed in three iterative
steps. The first iteration is used to estimate the noise spectrum; the second
iteration is used to identify weak emission lines and bad pixels and the third
and final gives the best-fit parameters. The spectrum in bad pixels and in
regions of line emission is replaced by the best-fit stellar spectrum. This step
is especially important for the Balmer absorption indices, which overlap regions
of nebular emission. Whether the higher-order Balmer lines are masked or not
does not affect our results \citep[$\mathrm{H}\epsilon$ and bluer;][]{barone+2020}.

Once the emission-line corrected spectra are determined, each index is measured after
convolving the spectrum with a Gaussian, so that the final spectral resolution matches
the spectral resolution of the relevant index. For more details on the fitting and
measurement procedure refer to \citet{scott+2017}.

In order to guarantee an acceptable precision, we bin the slit spectra out
from the central spaxel to guarantee a $\mathrm{SNR}=10 \, \mathrm{pixel}^{-1}$. Firstly
we fit the individual spectra, to estimate their SNR. We then fold the slit about the
central spaxel, ranking the spectra by their distance to the centre. Starting from the
central spaxel, we create spatial bins by summing adjacent spaxels until the
target SNR is met. For galaxies with obvious contamination, we consider only the half of
the spectrum away from the companion or interloper object.
Two example fits are shown in \reffig{f.da.fitsnrall}.

For our analysis, we focus on two indices: \hdeltaa and \fe4383. There are two
reasons for this choice: firstly, it represents a minimal index set that
is able to break the age-metallicity degeneracy, at least for the age range
relevant to PSB galaxies. In particular, \hdeltaa has a local maximum for a
$\approx 0.3 \,\mathrm{Gyr}$-old simple stellar population
\citep{worthey+ottaviani1997, kauffmann+2003a}, so that it is not possible to
invert the age-\hdeltaa function using \hdeltaa alone. However, for the ages and
metallicities relevant to this work, adding \fe4383 allows to break this age
degeneracy (see \S\ref{s.2ssp.ss.benchmark}).
Secondly, these indices ensure uniform coverage across
the largest possible sample, whereas indices defined at redder wavelengths drop
out of the LEGA-C observed range with increasing redshift.\\

Each index measurement has a measurement uncertainty, derived from the residuals of
the spaxel spectrum with respect to the best-fit spectrum. We find that these values
are underestimated and we derive an upscaling factor as follows.
Given the relatively high SNR in the central spaxels, we often have two measurements
at a given distance from the centre, one for each side of the slit. We assume that
galaxies are symmetric about their centre, so we can use the difference between the
two measurements to rescale the formal uncertainties on our measurements. Comparing the
measurements from either side of the galaxies, we find no systematic offset, but the
standard deviation is larger than the formal uncertainties. We rescale the formal
uncertainties by a factor that depends on the Lick index being considered and on the
value of the SNR. Given the SNR depends strongly on the distance from the centre of
each galaxy, our SNR rescaling factors are effectively a function of radius.
For \hdeltaa, the factor ranges from 1 (at the highest SNR) to 3 (for $10 < SNR < 15
\,\mathrm{pixel}^{-1}$).
For \fe4383, the factor ranges from 1 to 2.5.
Similar results were obtained following the method of \citet{straatman+2018},
i.e. using repeat observations of 61 galaxies to assess the random uncertainties
on the Lick index measurements. The main difference with our method is that using
repeat observations tends to overestimate the uncertainty for PSB galaxies, which
have systematically stronger \hdeltaa absorption compared to the sample of repeat
observations (only one PSB galaxy has two observations).

\begin{figure}
  \includegraphics[type=pdf,ext=.pdf,read=.pdf,width=1.0\columnwidth]{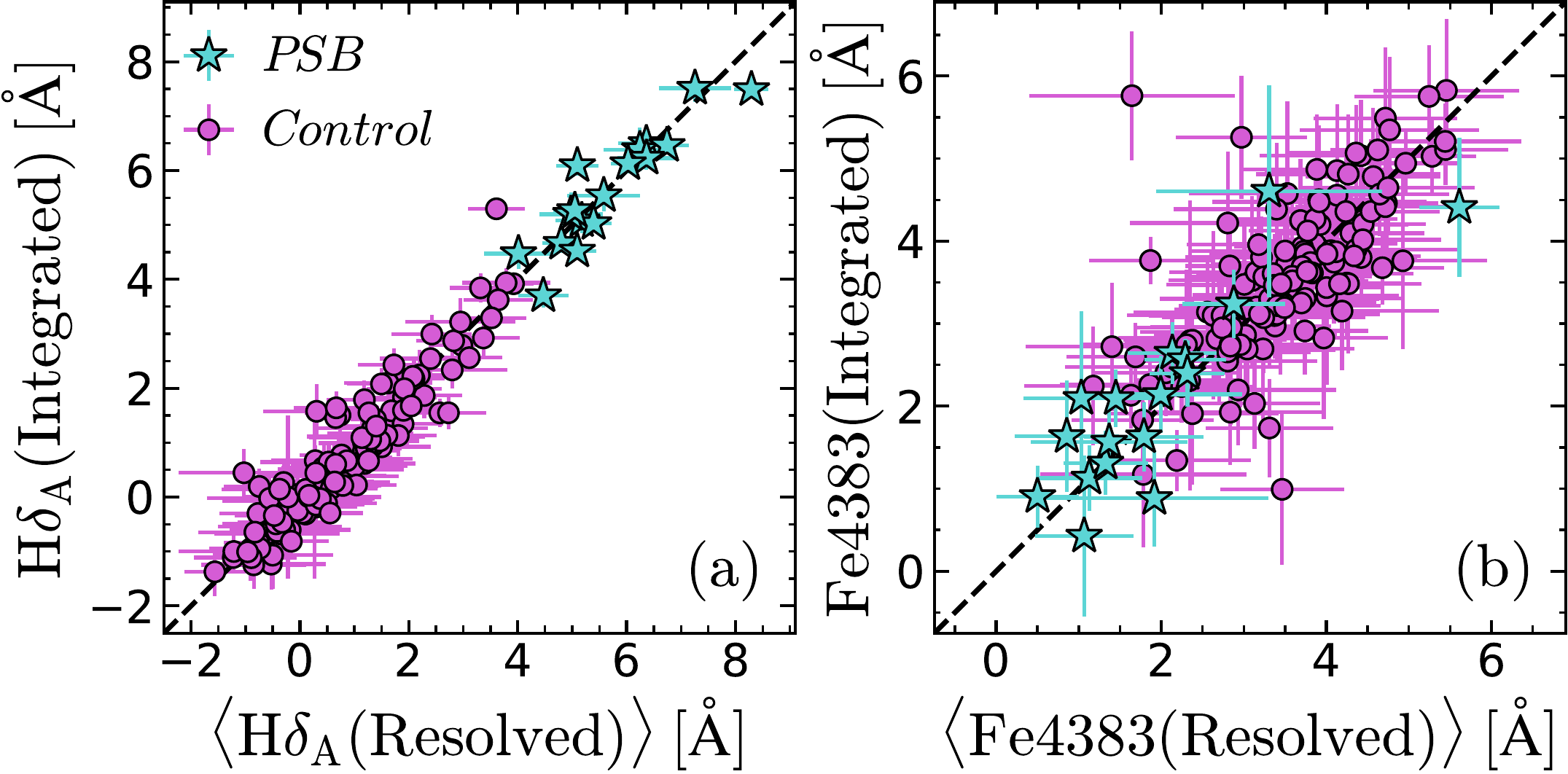}
  {\phantomsubcaption\label{f.da.index.comp.a}
   \phantomsubcaption\label{f.da.index.comp.b}}
  \caption{We find very good agreement between the median of the spatial
  measurements and the measurement on the integrated slit profile, for both
  \hdeltaa and \fe4383 (panels~\subref{f.da.index.comp.a} and
  \subref{f.da.index.comp.b} respectively). The cyan stars are PSB galaxies, the
  magenta circles are the control sample of quiescent galaxies (the errorbars have
  been rescaled). The best-fit
  relations have slopes $0.96\pm0.03$ (for \hdeltaa) and $1.03\pm0.08$ (for
  \fe4383), consistent with unity. The observed scatter about the best-fit
  relations are $\Delta = 0.26$ and $\Delta = 0.52$ respectively, consistent
  with the measurement uncertainties (after rejecting $3-\sigma$ outliers, the
  reduced $\chi^2$ value is 1.01 in both cases).
  }\label{f.da.index.comp}
\end{figure}

In \reffig{f.da.index.comp} we compare the value of the integrated Lick indices
from \citet{straatman+2018} to the (unweighted) median value for our resolved
measurements.
We show separately the PSB and control sample as cyan stars and magenta circles, but
we fit a single relation to both sets, and find excellent agreement between the
two measurements. For \hdeltaa we find a
best-fit linear slope of $0.96 \pm 0.03$ and a root-mean square residual along
the y-axis of $0.33 \, \text{\AA}$ (panel~\subref{f.da.index.comp.a}). Thus the
best-fit relation is statistically consistent with the identity. As for the scatter,
if we assume that the precision of the two determinations is the same and that
there was no intrinsic scatter (due e.g. to systematic errors), we can estimate
the average measurement uncertainty as $0.26/\sqrt{2} = 0.18 \, \text{\AA}$. 
\textit{Mutatis mutandis,} similar considerations apply to \fe4383; here the
best-fit linear relation has slope $1.03 \pm 0.08$ and observed scatter
$0.51$ (panel~\subref{f.da.index.comp.b}). In principle, a comparison between
the unweighted median and the integrated indices is biased, because the latter
are \textit{de facto} flux-weighted. However, if we repeat the above comparison
after replacing the unweighted median indices with the inverse-error weighted
indices, the results are statistically consistent with what we have reported for
the unweighted median (except for the best-fit slope of the \hdeltaa relation,
which goes from $0.96\pm0.03$ to $1.09\pm0.02$). This consistency is probably
due to the mix of PSB and non-PSB galaxies, because, as we will argue in the
next section, these two sets have different radial properties. These properties
are likely to impart opposite biases on the unweighted median indices compared
to the integrated indices.
Galaxy 107643~M4 is the most prominent outlier in \reffig{f.da.index.comp} (more
than three standard deviations), but it has a relatively bright interloper that
might affect the integrated spectrum. This galaxy is part of the control sample, and
its inclusion or removal does not change the outcome of our analysis.

When we repeat our analysis with the ELODIE stellar template library, we find
excellent agreement in the average value of the indices: considering only spectra
with $\mathrm{SNR} \geq 10 \, \mathrm{pixel}^{-1}$, we find no mean offset in either
\hdeltaa and \fe4383 ($\Delta \mathrm{H\delta_A} = -0.001 \pm 0.002 \, \text{\AA}$
and $\Delta \mathrm{Fe4383} = 0.012 \pm 0.009 \, \text{\AA}$).
\section{Results}\label{s.r}

\begin{figure}
  \includegraphics[type=pdf,ext=.pdf,read=.pdf,width=1.0\columnwidth]{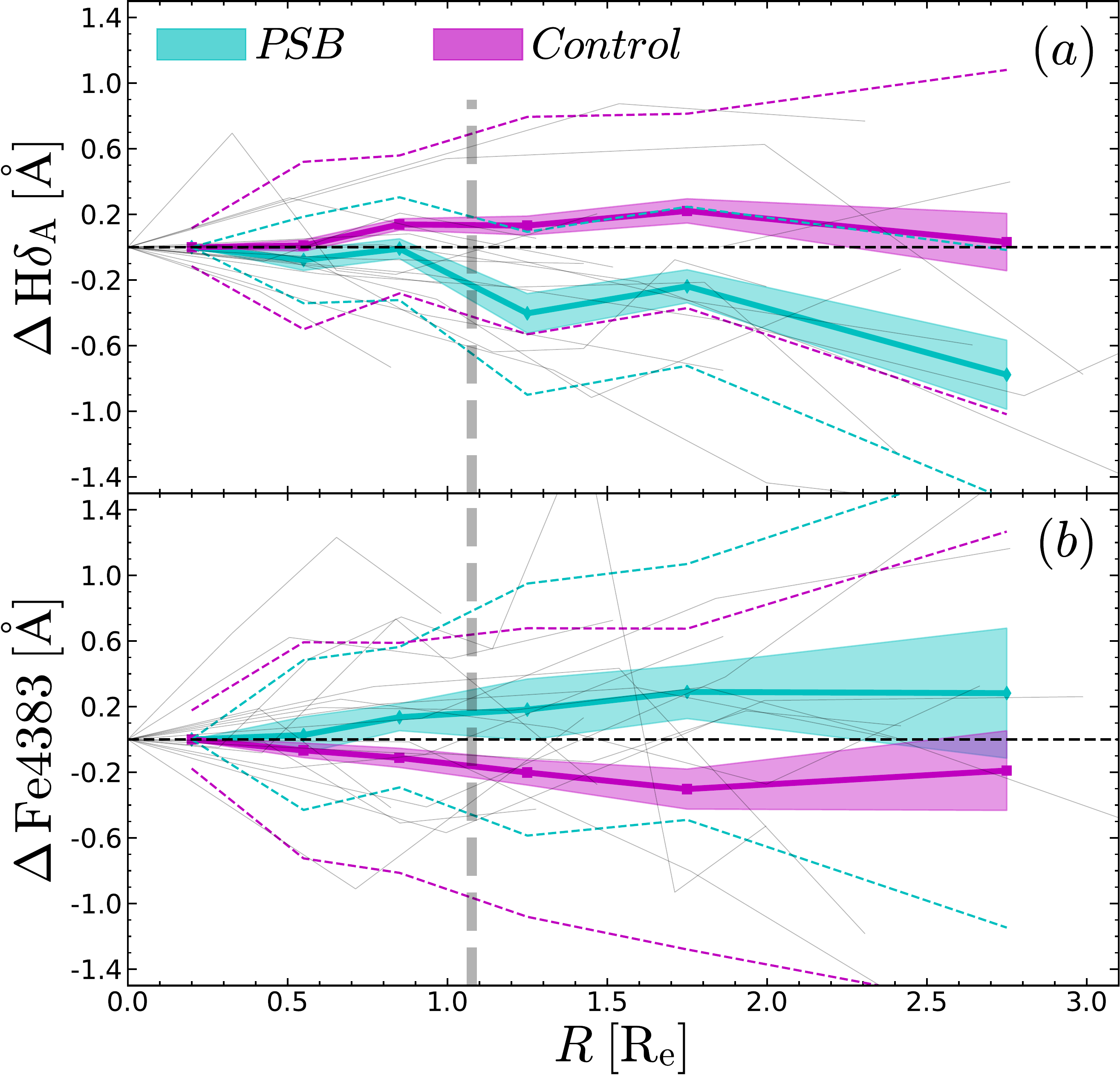}
  {\phantomsubcaption\label{f.r.indrad.a}
   \phantomsubcaption\label{f.r.indrad.b}}
  \caption{Unlike control galaxies (magenta), post-starburst galaxies (cyan) show
  decreasing \hdeltaa and flat \fe4383, clear signatures of a central starburst.
  The solid lines trace the running median of our measurements, the uncertainties
  about the median are enclosed by the shaded regions, whereas the coloured dashed
  lines enclose the 16\textsuperscript{th} and 84\textsuperscript{th} percentiles
  of the data. The vertical dashed lines marks the $\sigma$-equivalent of the
  median seeing. Thin grey lines trace the profile of individual PSB galaxies,
  showing that besides the average trend, individual galaxies present a range of
  radial profiles (individual control galaxies are not shown).
  The observed PSB trends are highly significant: in the most conservative
  estimate, the probability of a false-positive is $P=10^{-4}$.}\label{f.r.indrad}
\end{figure}

In \reffig{f.r.indrad.a} we show the average radial trends of \hdeltaa relative
to the central value, for both PSB galaxies (cyan) and the control sample
(magenta). The radial profiles of individual galaxies have been binned in
$R_e$, with the lines tracing the (moving) inverse-variance weighted median.
The uncertainty on the median is encompassed by the shaded region (estimated
as the semi-difference between the inverse-variance-weighted
16\textsuperscript{th} and 84\textsuperscript{th} percentiles,
divided by the square root of the number of measurements in each bin). The
dashed lines enclose the 16\textsuperscript{th} and 84\textsuperscript{th}
percentiles of the distribution.
The control sample of quiescent, non-PSB galaxies and the PSB sample have
opposite radial trends: the control sample has a weak positive \hdeltaa gradient.
In contrast, PSB galaxies have on average decreasing \hdeltaa profiles, i.e. the
\hdeltaa index is highest in the central regions and lowest in the outskirts.
For the \fe4383 index (\reffig{f.r.indrad.b}), we find that PSB galaxies
have a radially-increasing profile, whereas the control sample has decreasing
\fe4383 with radius.
Similar results are obtained for other empirical spectral indices, which we do
not show for brevity: for example, PSB galaxies have decreasing \hgammaa and
flat \dn4000, whereas control galaxies have increasing \hgammaa and decreasing
\dn4000 (see Appendix~\ref{app.otherinds}).

If we assume that PSB galaxies have flat \hdeltaa gradients, we can calculate
the probability of measuring by chance a negative gradient as follows: for each
radial measurement, we take the distance between the median \hdeltaa and zero
(the value expected from a flat gradient; this is equal to
$\Delta \; \mathrm{H\delta_A}$). We then divide this distance by the uncertainty
on $\Delta \; \mathrm{H\delta_A}$, and calculate the resulting one-tailed
probability of a value exceeding the measurement (we assume a Gaussian
distribution).
The number of independent radial measurements in the stacked profiles is
difficult to calculate, because each galaxy has different size and slightly
different seeing. We therefore provide the results for the most conservative
case only, i.e. assuming only two independent radial measurements. Assuming
PSB galaxies have flat \hdeltaa radial
profiles, the probability of finding by chance a negative gradient is
$P = 10^{-4}$. For the \fe4383 gradients, using the same assumptions we get
$P=0.04$. Bootstrapping 75\% of our data yields $P = 0.06$ and $P=0.04$
respectively. Similarly, the probabilities that the control sample has
flat profiles for \hdeltaa and \fe4383 are $P = 0.02$ and $P=0.07$ respectively
(bootstrapping yields $P=10^{-3}$ and $P=10^{-4}$).
Combining these opposite trends naturally leads to even smaller probabilities
that PSB and control galaxies have the same profiles: $P=10^{-5}$ and $P=0.01$
respectively, for \hdeltaa and \fe4383 (bootstrapping yields $P=0.03$ and
$P=0.01$). Assuming three independent radial bins yields $P$-values that are
$5-10$ times smaller.

\subsection{Caveats}\label{s.r.ss.caveats}

The results are qualitatively unchanged if we measure the distance along the
slit in physical units; however, since physical units may compound radial trends
\textit{within} galaxies with size trends \textit{between} galaxies, they are not
considered here.

We remark that the extent of our measurements is comparable to the FWHM of
the atmospheric seeing (the vertical dashed line in \reffig{f.r.indrad} is the
seeing equivalent $\sigma$, defined as FWHM/2.355, see \S\ref{s.da.ss.samp.sel}).
The value shown is the median seeing for the 17 PSB galaxies, expressed in units
of the galaxy \re, and falls approximately at one \re. For this reason, the
gradients measured here are much flatter than the intrinsic gradients
(\S\ref{s.2ssp}), in agreement with what is observed for PSB galaxies at
$z\approx0.1$ \citep{pracy+2013}. The fact that the seeing is comparable to
the median effective \re of our PSB sample might explain the change in slope
around $R \approx 2 \, \mathrm{R_e}$, but we cannot exclude the presence of
a size-dependent bias for this bin (only 13/17 measurements for the PSB sample,
only 36/141 measurements for the control sample).

Due to the small sample size, we are unable to study the relation between
inverse gradients and other galaxy properties. However, we find that the
trend is qualitatively unchanged if we consider mergers and non-mergers
separately, indicating that our results are not driven by prominent morphologic
asymmetries.
The same is true if we split the PSB sample in two at the median value of the
S{\'e}rsic index ($0.6 \leq n \leq 6$; the median is $2.9$), or at the median
value of the apparent effective radius ($0.2 \leq R_e \leq 0.7$; the median is
$0.3 \, \mathrm{arcsec}$), or at the median value of stellar mass
($10^{10.3} \leq M_\star \leq 10^{11.2}$; the median value is $10^{10.68} \,
\mathrm{M_\odot}$), or at the median value of the axis ratio ($0.21 \leq q \leq
0.94$; the median value is $0.64$).
Finally, we repeat the analysis limiting the selection to PSB classified as
central or isolated only: the trends are again qualitatively unchanged, ruling out
that our results are due to environment
effects on satellite galaxies.

In \S\ref{s.da.ss.samp.sel}, in order to ensure that the study is not limited
by the size of the control sample, we selected control galaxies to have the
same mass range as the PSB sample. Admittedly, a better choice would be to
select control galaxies having the same mass \textit{distribution} as the PSB
sample, because the strength of stellar population gradients of passive galaxies
depends on stellar mass and velocity dispersion \citep[e.g.][]{
martin-navarro+2018, zibetti+2020}.
However, the quiescent, non-PSB galaxies that make up the control sample have on
average lower SNR and larger measurement uncertainties than PSB galaxies of the
same mass. For this reason, imposing the same mass distribution between the PSB
and control samples results in too few control galaxies (25) with too large
measurement uncertainties to constrain the sample properties. Nevertheless, we
find that the mass-matched control sample is statistically consistent with the
actual control sample used in this work. To further test the effect of
mass-dependent bias in the observed gradients, we split the control sample in
two subsets at the value of its median stellar mass $10^{10.82} \,
\mathrm{M_\odot}$. We find that both subsets have gradients that are
statistically consistent with the control sample, but the most-massive half of
the sample has steeper \fe4383 gradient than the least-massive half, in
qualitative agreement with observations of local galaxies \citetext{e.g.
\citealp{martin-navarro+2018} find that the most massive elliptical galaxies
have steeper radial metallicity gradients}. Using either half of the control
sample would not change the nature of our results.

Incidentally, the fact that PSB galaxies have opposite radial trends compared to
the general population suggests that our results are unlikely to arise from bias due
to decreasing SNR with radius.

\section{A two-SSP toy model}\label{s.2ssp}

\begin{figure*}
  \includegraphics[type=pdf,ext=.pdf,read=.pdf,width=1.0\textwidth]{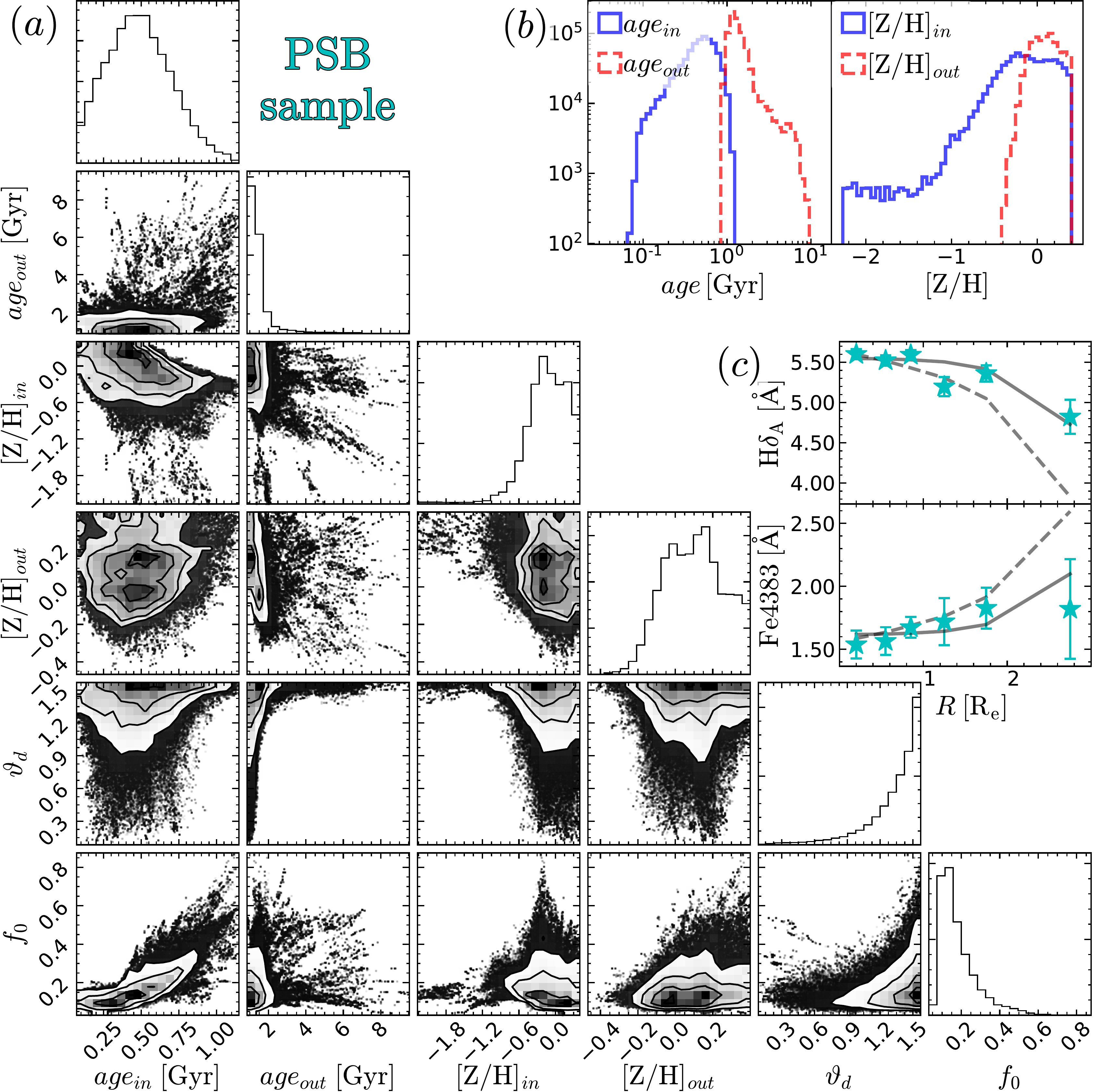}
  {\phantomsubcaption\label{f.r.psbmodel.a}
   \phantomsubcaption\label{f.r.psbmodel.b}
   \phantomsubcaption\label{f.r.psbmodel.c}}
  \caption{Our two-SSP toy model requires inverse age gradients in PSB
  galaxies (Panel~\subref{f.r.psbmodel.a}). The model uses two SSPs with different
  spatial distribution to reproduce the median observed gradients in both
  \hdeltaa and \fe4383 (Panel~\subref{f.r.psbmodel.c}). The dashed/solid lines show
  the model prediction before/after seeing convolution, and the cyan stars trace
  the median index values for the PSB sample. The age of the inner and outer SSP
  are clearly different: the inner SSP (solid blue histogram in
  Panel~\subref{f.r.psbmodel.b}) is clearly younger than the outer SSP (dashed red
  histogram). In contrast, SSP metallicities are consistent within the
  uncertainties. The corner diagram shows the marginalised probability for the
  model; the strongest correlation is between $f_0$ (tracing the burst fraction)
  and $age_{in}$ (tracing the burst age).
  }\label{f.r.psbmodel}
\end{figure*}
To interpret the observed trends, we implement a six-parameter model to
predict the stacked measurements of \reffig{f.r.indrad}. As a light profile, we
use a one-dimensional S{\'e}rsic model, where the spectrum at each radius is the
superposition of two simple stellar populations (SSP; i.e. each population has
uniform age and metallicity). As SSP spectra we take the MILES models
\citep{vazdekis+2010, vazdekis+2015}, using BaSTI isochrones
\citep{pietrinferni+2004, pietrinferni+2006},
solar $[\alpha/\mathrm{Fe}]$ and Chabrier IMF \citep{chabrier2003}. The
resulting grid
of 636 spectra spans $-2.27 < [\mathrm{Z/H}] < 0.40$ and $0.03 < age < 14.00
\, \mathrm{Gyr}$ \citetext{replacing BaSTI with Padova isochrones from
\citealp{bertelli+2009} yields qualitatively consistent results}.
Our model superimposes two SSPs, representing the central stars and the stars
in the outskirts of the galaxy, labelled respectively
``in'' and ``out'' (the corresponding SSP parameters are $age_{in}$,
$\mathrm{[Z/H]}_{in}$, $age_{out}$ and $\mathrm{[Z/H]}_{out}$). The mass
fraction of the ``in'' SSP to the total is given, at each radius $R$, by:
\begin{equation}\label{eq.r.frac}
    f(R) = f_0 \dfrac{e^{\frac{R_m-R}{R_d}}-1}{e^{\frac{R_m}{R_d}}-1}
\end{equation}
where $R$ is expressed in units of $R_e$, and $R_m = 6 \, \mathrm{R_e}$ is an
arbitrary radius that is ``large'' relative to the extent of our measurements.
$f$ is a declining exponential function, scaled so that the central value is $f(0) = f_0$
and downshifted so that $f(R_m)=0$. This choice is motivated as follows.
Firstly, stellar generations tend to form superimposed exponential discs
\citetext{\citealp{poci+2019}, \citealp{buck+2019} - and the ratio of two
exponentials is also exponential}.
Secondly, there is evidence that PSB galaxies host rotation-supported discs
\citep{hunt+2018}.
The parameter $R_d$ specifies the concentration of the central SSP: for any
non-negative value, $R_d$ is the exponential scale radius of $f(R)$, in the
sense that $\partial_R f = -f/R_d + const.$ (smaller values of $R_d$ correspond
to more concentrated central populations). As 
$R_d \rightarrow \infty$, we have $f(R) \rightarrow (1 - R/R_m)$: in other
words, using Eq.~(\ref{eq.r.frac}) to express $f(R)$ includes both a
physically-motivated exponentially-declining fraction, and a linear mixing
fraction which represents the simplest uninformed guess. In practice we
implement the infinite range in $R_d$ by parametrising this scale radius as
$\tan \vartheta_d$, with $0 \leq \vartheta_d \leq \pi/2$. Thus the fraction $f$
requires two additional parameters: $f_0$ and $\vartheta_d$. For our purposes,
$f_0$ and $\vartheta_d$ are just nuisance parameters: given (i) our spatial
resolution and (ii) the use of a stack analysis, we cannot meaningfully
constrain the \textit{structure} of PSB galaxies, but just the sign of radial
gradients of stellar age and metallicity.
The S{\'e}rsic profile has arbitrary central surface brightness and $R_e$, but
the S{\'e}rsic index is fixed at $n=2.4$, the median value for the PSB sample.
The model is convolved with a Gaussian PSF with $\sigma = 1 \, R_e$
(see \reffig{f.r.indrad}). In summary,
our most general model has six free parameters, the age and metallicity for each
of the two SSPs, and two more parameters to specify the (monotonic) radial mixing
of these two SSPs. The likelihood of the data given these model parameters is
expressed as a multivariate Gaussian over the observed \hdeltaa and \fe4383
measurements.
We assume flat priors on all the model parameters, with the allowed range
equal to the physical range of each parameter:
$0.03 < age_{in}, age_{out} < 14.00 \, \mathrm{Gyr}$, $-2.27 \leq
\mathrm{[Z/H]}_{in},\; \mathrm{[Z/H]}_{out} \leq 0.40$, $0 \leq \vartheta_d \leq
\pi/2$ and $0 \leq f_0 \leq 1$. We combine the likelihood and priors to write
the posterior distribution (apart from the evidence), and we estimate the
model parameters by integrating the posterior distribution with the Markov Chain
Monte Carlo approach \citep{metropolis+1953}.

Constraining this six-parameter model using twelve measurements is problematic,
but our goal is not to infer an accurate value of the parameters. We use the
model as a benchmark, to assess the plausibility of less general sub-models,
obtained by constraining some of the six parameters from the general model, and we
show that any acceptable solution has properties that are inconsistent with the
properties of the control sample.

\subsection{The benchmark model for PSB galaxies}\label{s.2ssp.ss.benchmark}

The results for the most general PSB model are shown in \reffig{f.r.psbmodel}.
Panel \subref{f.r.psbmodel.a} shows the posterior distribution for the six model
parameters, marginalised over all possible sets of four and five parameters.
A summary of the model results is reported in Table~\ref{t.r.models}.
For ease of comparison, the posterior distributions of $age_{in}$ and
$age_{out}$ and of $\mathrm{[Z/H]}_{in}$ and $\mathrm{[Z/H]}_{out}$ are
reported also in panel~\subref{f.r.psbmodel.b} (using a logarithmic scale for age).
Panel~\subref{f.r.psbmodel.c} compares the measured Lick indices (cyan stars with
errorbars) to the prediction of the most likely model (i.e. the mode of the
posterior distribution); the dashed and solid grey lines trace respectively the
intrinsic profile and the seeing convolved profile.
Within one \re, the intrinsic \hdeltaa gradient for our stacked profile is
$\approx -0.76\pm0.03 \, \text{\AA}/R_e$, consistent with the median
value for local PSB galaxies \citetext{$-0.83\pm0.23 \, \text{\AA}/R_e$;
\citealp{chen+2019}. We considered both their ``central'' and ``ring-like'' PSBs to
calculate the median, consistent with our sample selection criteria that do not
differentiate between different PSB morphologies. The fit was performed using
weighted least squares optimization.}

The corner plot shows the well-known age-metallicity degeneracy \citep{
worthey1994}, for each of the
two SSPs independently. The Spearman rank correlation coefficient is
$\rho = -0.39$ for $age_{in}$ and $\mathrm{[Z/H]}_{in}$ and is $\rho = -0.52$
for $age_{out}$ and $\mathrm{[Z/H]}_{out}$ (panel~\subref{f.r.psbmodel.a}; all
$P$-values are zero owing to the large number of sample points).
Of all the possible parameter pairs, we find the strongest degeneracy between
$f_0$ and $age_{in}$ ($\rho = 0.84$); this can be interpreted as the degeneracy
between the burst fraction (governed by the central value $f_0$) and the burst
age \citep{serra+trager2007}.

With these degeneracies in mind, we can inspect panel \subref{f.r.psbmodel.b},
which reports the age and metallicity histograms of the two SSPs on the same
scale: here the central SSP (solid blue histogram) has both younger age and
lower metallicity than the outer SSP (red dashed histogram). For SSP age,
we find $age_{in} = 0.48^{+0.23}_{-0.21} \, \mathrm{Gyr}$ and
$age_{out} = 1.28^{+0.43}_{-0.20} \, \mathrm{Gyr}$ (here and in the following
the results quoted refers to the 50\textsuperscript{th} and the uncertainties
encompass the 16\textsuperscript{th} and 84\textsuperscript{th} percentiles of
the relevant posterior distribution).
For metallicity, the results are
$\mathrm{[Z/H]}_{in} = -0.13^{+0.33}_{-0.22}$ and
$\mathrm{[Z/H]}_{out} = 0.10^{+0.17}_{-0.06}$.
We find that metallicity is not well constrained, as expected from young SSPs.
However, while the difference in $\mathrm{[Z/H]}$ is not statistically
significant (within one standard deviation), the age difference is larger than
3.5 standard deviations: the probability $P$ that the two SSPs have the same age
is $P<2\times10^{-4}$ (this value assumes a Gaussian distribution and is the
most conservative result; using the marginalised posterior distribution we
obtain $P<1.3\times 10^{-6}$). The strong separation between the age of the two
SSPs is mostly due to the sharp cutoff in the posterior distribution of
$age_{out}$ below $\approx 0.9 \, \mathrm{Gyr}$ (red dashed age histogram in
panel \subref{f.r.psbmodel.b}). This strong cutoff may be surprising, because
\hdeltaa has a local maximum at $0.1-1\,\mathrm{Gyr}$ \citetext{the exact value
depends on metallicity, see e.g. \citealp{worthey+ottaviani1997} and
\citealp{kauffmann+2003a}, their fig.~2}, so that there is a strong degeneracy
between \hdeltaa and SSP age precisely where the model infers a cutoff in the
distribution. The solution to this apparent conundrum is in the value of \fe4383:
at fixed metallicity, \fe4383 increases with SSP age. For this reason, even
though $age_{out}$ younger than $0.9 \, \mathrm{Gyr}$ could indeed explain the
decreasing values of \hdeltaa with radius, it would also predict a radially
decreasing \fe4383, opposite to what is observed. In addition, even for the
highest metallicity, SSPs young enough to have $\mathrm{H\delta_A} \lesssim 4 \,
\text{\AA}$ have $\mathrm{Fe4383}<0\,\text{\AA}$, inconsistent with the
observations. Notice also that while $age_{in} < age_{out}$, we do not find
an ``old'' outer SSP: this is due to the use of a single SSP instead of an
extended SFH, as we show in \S\ref{s.2ssp.ss.3ssp}.
We conclude that this model, simple yet general, strongly prefers an age
gradient over a metallicity gradient to explain the median index profiles
observed in our PSB sample.

\subsection{Modelling the control sample}\label{s.2ssp.ss.control}

\begin{figure}
  \includegraphics[type=pdf,ext=.pdf,read=.pdf,width=1.0\columnwidth]{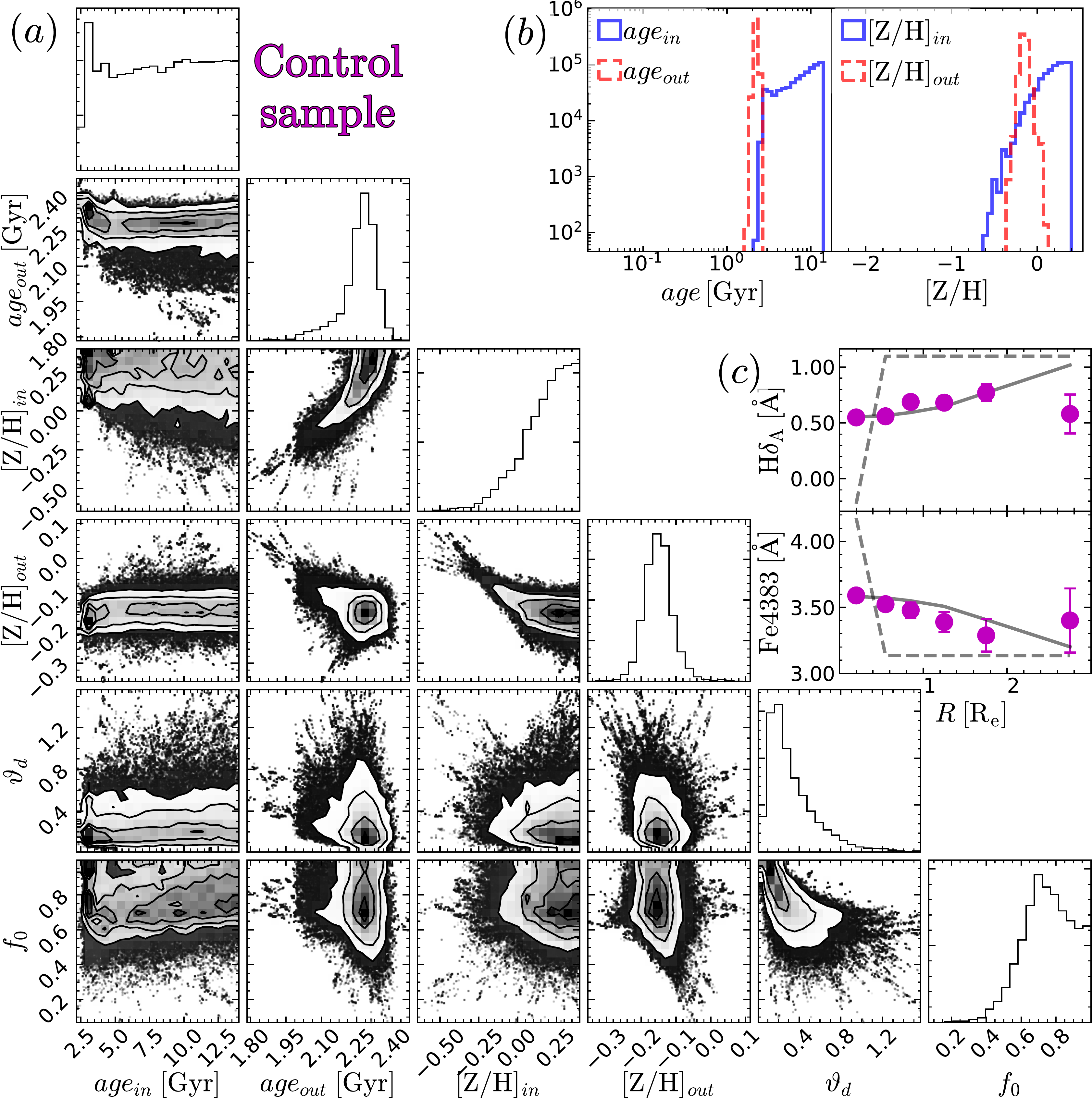}
  {\phantomsubcaption\label{f.r.conmodel.a}
   \phantomsubcaption\label{f.r.conmodel.b}
   \phantomsubcaption\label{f.r.conmodel.c}}
  \caption{Our two-SSP toy model requires inside-out age and/or metallicity gradients
  for the control sample of quiescent, non PSB galaxies
  (panels~\subref{f.r.conmodel.a} and \subref{f.r.conmodel.b}). The inferred
  model reproduces the median observed gradients in both \hdeltaa and \fe4383
  (panel~\subref{f.r.conmodel.c}; the dashed/solid lines show the model prediction
  before/after seeing convolution, and the magenta circles trace the median index
  values for our control sample). Even though the age of the inner SSP is poorly
  constrained, it is clear that $age_{in} > age_{out}$
  (panel~\subref{f.r.conmodel.b}): the inner SSP (solid blue histogram)
  is clearly older than the outer SSP (dashed red histogram, $P<3\times10^{-4}$).
  In contrast, even though $\mathrm{[Z/H]}_{in} > \mathrm{[Z/H]}_{out}$, the two
  SSP metallicities are consistent within the uncertainties ($P<0.07$). Notice
  the strong degeneracy between the parameters of the two SSPs.
  }\label{f.r.conmodel}
\end{figure}

We have already ruled out the possibility that the median gradient of the PSB
and of the control sample are the same. But could these different gradients
arise from similar stellar populations, observed at different ages? This question
is paramount to understanding whether (in an average sense) the control sample is
consistent with passive evolution of the PSB sample. To give an answer, we apply
our toy model to the median Lick profiles of the control sample. The results are
illustrated in \reffig{f.r.conmodel}, where the meaning of the symbols are the
same as in \reffig{f.r.psbmodel}.
The model predictions are compared to the data in panel~\subref{f.r.conmodel.c}:
within one \re, we find a \hdeltaa gradient of $\approx 1 \, \text{\AA}/R_e$, whereas
outside one \re the intrinsic index profile is flat. This behaviour results from
the combination of a relatively high central fraction (see the posterior
probability of $f_0$ in panel~\subref{f.r.conmodel.a}) and compact spatial
distribution (small $\vartheta_d$). This behaviour however may not be robust,
given the strong degeneracy between $f_0$ and $\vartheta_d$, which affects the
intrinsic radial gradients ($\rho=-0.76$). We also remark that the last radial
measurement appears to be an outlier: ignoring this point yields a less steep
gradient of $\approx 0.8 \, \text{\AA}/R_e$.

Examining the corner diagram (panel~\subref{f.r.conmodel.a}), we find again that
age and metallicity anticorrelate for each SSP ($\rho = -0.10$ and $-0.31$ for
the inner and outer
SSP, respectively). However, compared to the posterior probability distribution
for the PSB sample, for the control sample these degeneracies are significantly
smaller (in absolute value). On the other hand, we find a strong correlation of
the metallicity of the inner SSP ($\mathrm{[Z/H]}_{in}$) with both age and
metallicity of the outer SSP: we find $\rho=0.61$ for $age_{out}$ and
$\rho=-0.72$ for $\mathrm{[Z/H]}_{out}$. These strong correlations are likely due
to metallicity being the strongest driver of both \hdeltaa and \fe4383 for old
stellar populations; their presence highlights the need for comprehensive
modelling in order to interpret our data.

Panel~\subref{f.r.psbmodel.b} shows the age and metallicity histograms of the
two SSPs: unlike for PSB galaxies, here the central SSP (solid blue histogram)
has both older age and higher metallicity than the outer SSP.
We find $age_{in} = 8.3^{+3.9}_{-4.2} \, \mathrm{Gyr}$ whereas
$age_{out} = 2.27^{+0.08}_{-0.05} \, \mathrm{Gyr}$; for metallicity, the
results are $\mathrm{[Z/H]}_{in} = 0.21^{+0.13}_{-0.19}$ and
$\mathrm{[Z/H]}_{out} = -0.15^{+0.04}_{-0.04}$. For control galaxies, the
probability that the two ages are the same is $P<3\times10^{-4}$ (using the
joint posterior probability distribution of $age_{in}$ and $age_{out}$)\footnote{
In this case, assuming a Gaussian probability with standard deviation equal to
the standard deviation of the posterior yields $P<0.1$, but this estimate is too
conservative: the large difference in probability between the two estimates
arises from the shape of the
posterior probability distribution of $age_{in}$, that has a sharp cutoff below
$\approx 2.5 \, \mathrm{Gyr}$ (\reffig{f.r.conmodel.a}).} It is unclear whether
and to what extent this sharp cutoff might be caused by our stacking analysis.
In fact, on one hand the youngest galaxies are (on average) the smallest, which
enhances their contribution to the innermost radial bins and biases $age_{in}$ to
younger values; on the other hand, however, the oldest galaxies, despite being on
average the largest, tend to have the steepest light profiles, which also enhances
their contribution to the innermost bins. Disentangling these two competing
effects is however beyond the scope of this paper.
From \reffig{f.r.conmodel.b}, we also notice that SSP age
for both the inner and outer SSPs is larger than in PSB galaxies. This might be
due either to the different mass distribution of the two samples, but also to
problems inherent with the stacking analysis. For example, if the youngest PSB
galaxies are also the most compact, their PSF would be larger than the median
PSF (when expressed in units of $R_e$), causing more contamination between the
inner and outer SSP. In light of this ambiguity, we do not overinterpret the
observed age difference between control galaxies and the outer SSP of PSB
galaxies.
For metallicity, we find $\mathrm{[Z/H]}_{in} > \mathrm{[Z/H]}_{out}$, but
this result is not statistically significant (the probability that the two SSPs
have the same metallicity is $P<0.07$).

At face value, however, we find that modelling the control sample requires the
central SSP to be older and more metal rich than the outer SSP, as observed in
most local quiescent galaxies \citep[e.g.][]{mcdermid+2015, zibetti+2020,
ferreras+2019} and at variance with the model for PSB galaxies.

\subsection{Constrained PSB models}\label{s.2ssp.ss.constrained}

\begin{table*}
  \begin{center}
  \setlength{\tabcolsep}{2pt}
  \caption{Summary of the two-SSPs models. PSB galaxies are best described by an
  inverse stellar population structure, i.e. centre younger and/or lower
  metallicity than the outskirts. Imposing an inside-out age structure on PSB
  galaxies yields a poor fit (reduced $\chi^2 \approx 6$).
  On the contrary, control galaxies are best described by an inside-out stellar
  population structure.}\label{t.r.models}
  \begin{tabular}[t]{cccccccccc}
  \hline
  Model Name & Constraints & $age_{in}$ & $age_{out}$ & $\mathrm{[Z/H]}_{in}$ & $\mathrm{[Z/H]}_{out}$ & $\vartheta_d$ & $f_0$ & $\chi^2_\nu$ & Structure \\
      &     & $\mathrm{Gyr}$ & $\mathrm{Gyr}$ &     &     &     &     &    & \\
  (1) & (2) & (3) & (4) & (5) & (6) & (7) & (8) & (9) & (10) \\
  \hline
   \specialcell{PSB\\(benchmark)} & none & $0.48^{+0.23}_{-0.21}$ & $1.28^{+0.43}_{-0.20}$ & $-0.13^{+0.33}_{-0.22}$ & $ 0.10^{+0.17}_{-0.06}$ & $1.41^{+0.13}_{-0.29}$ & $0.15^{+0.11}_{-0.05}$ & 1.9 & inverse \\
   \specialcell{Control\\(benchmark)} & none & $8.3^{+3.9}_{-4.2}$    & $2.27^{+0.05}_{-0.08}$ & $ 0.21^{+0.13}_{-0.19}$ & $-0.15^{+0.04}_{-0.04}$ & $0.26^{+0.28}_{-0.13}$ & $0.75^{+0.15}_{-0.14}$ & 3.5 & inside-out \\
   \specialcell{PSB inside-out\\age} & $age_{in}\geq age_{out}$ & $3.95^{+7.70}_{-3.10}$ & $0.85^{+0.02}_{-6.28}$ & $ 0.08^{+0.18}_{-0.69}$ & $ 0.15^{+0.07}_{-0.10}$ & $1.20^{+0.34}_{-0.87}$ & $0.40^{+0.59}_{-0.34}$ & 6.0 & \specialcell{inside-out (age)\\inverse (metallicity)}\\
   \specialcell{PSB inside-out\\metallicity}& $\mathrm{[Z/H]}_{in}\geq\mathrm{[Z/H]}_{out}$ & $0.68^{+0.12}_{-0.11}$ & $9.84^{+2.99}_{-6.97}$ & $ 0.18^{+0.15}_{-0.08}$ & $ 0.06^{+0.13}_{-0.15}$ & $1.51^{+0.04}_{-0.08}$ & $0.30^{+0.14}_{-0.08}$ & 2.0 & \specialcell{inverse (age)\\inside-out (metallicity)} \\
  \hline
  \end{tabular}
  (1) Name of the model as introduced in the main text. (2) Additional
  constraints on the age and metallicity of the two SSPs. (3) Inferred age of
  the central SSPs (here and in the following, we quote the median value of the
  marginalised posterior distribution; the uncertainties refer to the
  16\textsuperscript{th} and 84\textsuperscript{th} percentile of the probability).
  (4) Inferred age of the outer SSP. (5) Inferred metallicity of the central SSP.
  (6) Inferred metallicity of the outer SSP. (7) Inferred value of the
  concentration parameter. (8) Inferred value of the mass fraction of the central
  SSP in the central pixel. (9) $\chi^2$ per degree of freedom. (10) Description
  of the model outcome: ``inverse'' refers to positive radial gradients in age
  and/or metallicity, ``inside-out'' refers to negative radial gradients in age
  and/or metallicity.
  \end{center}
\end{table*}

Could an inside-out SSP gradient reproduce the observed radial trends of \hdeltaa
and \fe4383 for PSB galaxies? We have already shown that the benchmark model
points to an inverse age structure for PSB galaxies, therefore we know that an
inside-out structure is less likely. However, our two-SSP model has only six
degrees of freedom, therefore its predictive power is modest. For this reason,
it is important to evaluate directly how much worse an inside-out model would be
relative to the benchmark model. To address this
question, we create two more PSB models, identical to the benchmark model, but
constrained to have inside-out age or metallicity gradients.
The first model has free age but $\mathrm{[Z/H]}_{in} > \mathrm{[Z/H]}_{out}$
(PSB inside-out metallicity model, Table~\ref{t.r.models}). For this model, we
find that $age_{in} < age_{out}$, consistent with the benchmark model; the fact
that - by construction - $\mathrm{[Z/H]}_{in} > \mathrm{[Z/H]}_{out}$ yields a
marginally higher $\chi^2_\nu = 2.0$ compared to the benchmark $\chi^2_\nu =
1.9$.
The second model has free metallicity, but $age_{in} > age_{out}$ (PSB
inside-out age model, Table~\ref{t.r.models}). The best-fit parameters for this
model predict flat \hdeltaa and \fe4383 radial profiles, inconsistent with
observations. Quantitatively, the reduced $\chi^2_\nu \approx 6$ is larger than
the value for the fiducial model ($\chi^2_\nu = 1.9$). We conclude that an
inside-out age structure is inconsistent with the radial variations of \hdeltaa
and \fe4383 observed for the stacked PSB galaxies.
These two constrained models suggest that, while inside-out age gradients are
ruled out for the PSB sample, both inside-out and inverse metallicity
gradients are consistent with observations \citep[cf.][]{cresci+2010,
schonrich+mcmillan2017}.

\subsection{Effect of extended star-formation history}\label{s.2ssp.ss.3ssp}

While an SSP is a good model for a starburst (where the spread in stellar age is
narrow by definition), stellar populations are known to have extended
star-formation histories (SFH). To what extent the different properties of SSPs
and more realistic stellar population might bias our results? To address this
question, we implemented a three-SSP model. For
brevity, we refer to these SSPs by increasing roman numerals I-III. In order to
preserve some predictive power, we want the smallest possible number of free
parameters. For this reason, the three SSPs have different ages
($age_\mathrm{I}$, $age_\mathrm{II}$ and $age_\mathrm{III}$) but equal
metallicity $\mathrm{[Z/H]}$. Even though this restriction is not realistic, it
reflects the fact that, with our two indices, we do not find strong metallicity
differences within the PSB sample (right panel of \reffig{f.r.indrad.b}).
For the radial variation, we use the same parametrisation introduced for the
benchmark model (Eq.~\ref{eq.r.frac}; again, like we did for the benchmark
model, we parametrise $R_d$ as $\tan \vartheta_d$, with $0 \leq \vartheta_d \leq
\pi/2$). Using the same
parametrisation means that SSP~I can be safely interpreted as the inner SSP from
the benchmark model, whereas the outer SSP from the benchmark model corresponds
here to the superposition of SSP~II and SSP~III: this superposition is the
zero-order approximation for an extended SFH.
At any radius $R$, the value of the mass fractions are $f_\mathrm{I}(R)$,
$f_\mathrm{II}(R)$ and $f_\mathrm{III}(R)$, that we parametrise using two
variables $0\leq \vartheta \leq \pi/2$ and $0\leq \varphi \leq \pi/2$:
\begin{equation}
  \begin{cases}
    f_\mathrm{I}(R)   \equiv &      \cos^2 (\vartheta) \, f(R) \\
    f_\mathrm{II}(R)  \equiv & [1 - \cos^2 (\vartheta) \, f(R) ] \, \cos^2 (\varphi) \\
    f_\mathrm{III}(R) \equiv & [1 - \cos^2 (\vartheta) \, f(R) ] \, \sin^2 (\varphi)
  \end{cases}
\end{equation}
It can be easily verified that these functions express meaningful fractions,
because they satisfy both $f_\mathrm{I}(R) + f_\mathrm{II}(R) + f_\mathrm{III}(R)
= 1$ as well as $0\leq f_\mathrm{x}(R) \leq 1;\;
\forall x\in\{\mathrm{I,II,III}\}$. In particular, the angle $\vartheta$
expresses the fraction of SSP~I at $R=0$, via $f_\mathrm{I}(0)=\cos^2
(\vartheta)$, whereas the angle $\varphi$ expresses the fraction, relative to
the \emph{remaining} stellar population, of SSP~II (resp. SSP~III) via $\cos^2
(\varphi)$ (resp. $\sin^2 (\varphi)$). Notice that the fraction of SSP~II
decreases with increasing $\varphi$, whereas the fraction of SSP~III increases
with increasing $\varphi$. This model has an undesired symmetry, in that the
results are unchanged swapping $(age_\mathrm{II}, age_\mathrm{III}, \varphi)$
with $(age_\mathrm{III}, age_\mathrm{II}, \pi/2-\varphi)$, therefore we further
require $age_\mathrm{II} \leq age_\mathrm{III}$ by assigning zero probability to
non-complying models. This means that SSP~II is always younger than SSP~III,
therefore the age of the outer SSP (equal to the sum of SSP~II and SSP~III)
increases with increasing $\varphi$.
So far the model has seven free parameters: $age_\mathrm{I}$, $age_\mathrm{II}$,
$age_\mathrm{III}$, $\mathrm{[Z/H]}$, $\vartheta$, $\varphi$ and $\vartheta_d$.
To reduce this number, we further constrain $\vartheta$ and $\vartheta_d$ to
reproduce the corresponding optimal values from the benchmark model, i.e. we set
$\vartheta_d = 1.41$ and $\vartheta = \arccos \sqrt{f_0}$, with $f_0 = 0.15$
(first row of Table~\ref{t.r.models}).

\begin{figure}
  \includegraphics[type=pdf,ext=.pdf,read=.pdf,width=1.0\columnwidth]{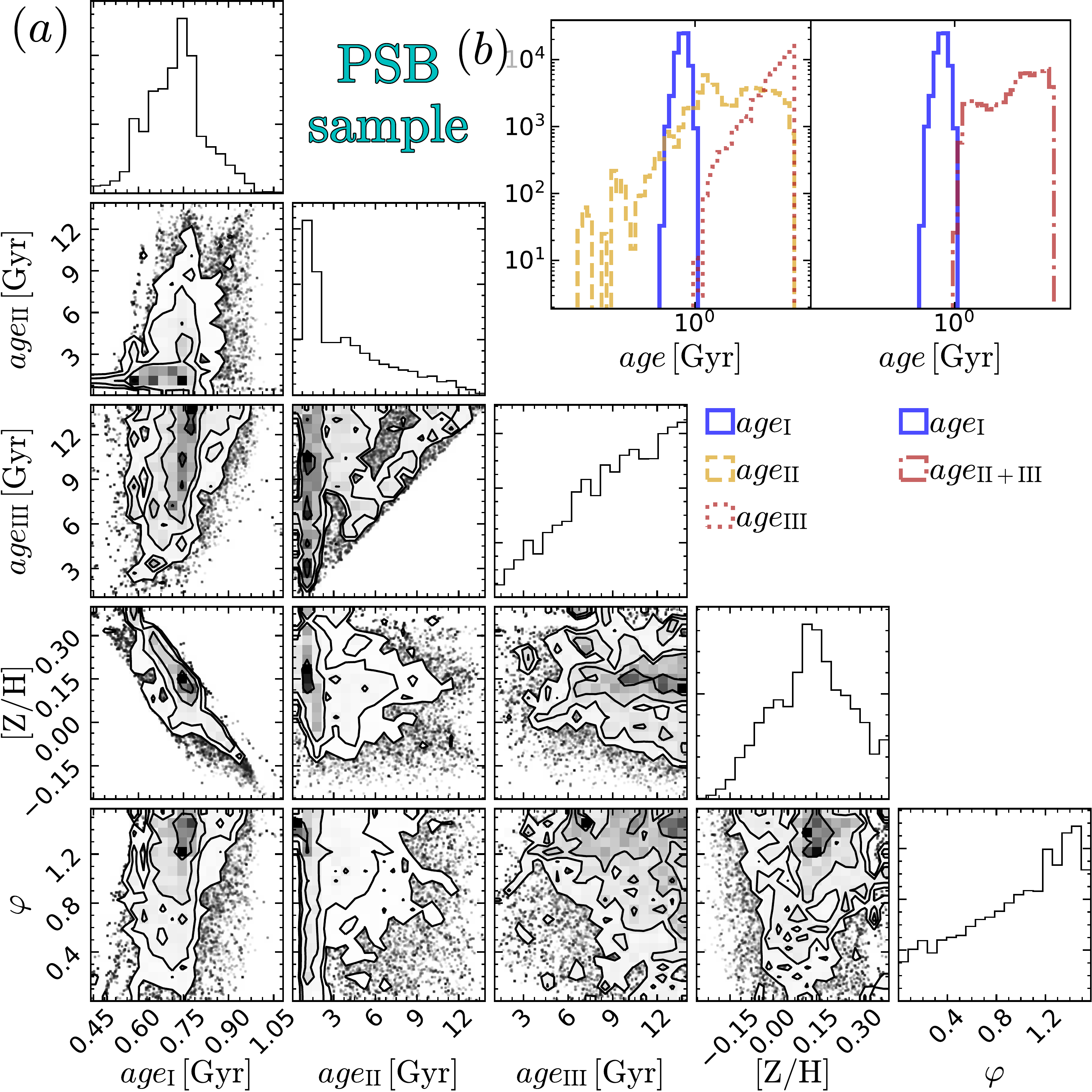}
  {\phantomsubcaption\label{f.r.thrage.a}
   \phantomsubcaption\label{f.r.thrage.b}}
  \caption{Assessing the effect of extended star-formation histories instead of
  SSPs for PSB galaxies. The model uses three SSPs: SSP~I corresponds to the
  inner stellar population, whereas the superposition of the young SSP~II and the
  old SSP~III are a first-order approximation to an extended star-formation
  history. The corner diagram (panel~\subref{f.r.thrage.a}) shows that the fraction
  of SSP~III stars ($\cos^2\,
  \varphi$) anti-correlates with  $age_\mathrm{II}$ (the age of the younger
  SSP~II) and correlates with $age_\mathrm{III}$ (the age of the older SSP~III).
  This behaviour reflects the fact that the indices considered here constrain
  only the average age of the outer stellar population: this value of the age can
  be attained with both a long or a short star-formation history.
  Panel~\subref{f.r.thrage.b}, left, shows the probability distribution of the
  three SSP ages: $age_\mathrm{I}$ (solid blue) is clearly younger than both
  $age_\mathrm{II}$ (dashed yellow) and $age_\mathrm{III}$ (dot-dashed red). Even
  though occasionally $age_\mathrm{II} \leq age_\mathrm{I}$, this occurs with
  relatively low probability $P<0.07$. More importantly, in the right panel
  we can see that the inner SSP is systematically younger than the outer SSP
  $P < 10^{-5}$.
  }\label{f.r.thrage}
\end{figure}

This model reproduces the observed index profiles as well as the benchmark model
($\chi^2_\nu = 10^{1000}$); the marginalised posterior probability is
illustrated by the corner diagram in Fig.~\ref{f.r.thrage.a}.
We find the usual age-metallicity degeneracy, between each SSP age and
$\mathrm{[Z/H]}$ (the Spearman rank correlation coefficients for
$age_\mathrm{I}$, $age_\mathrm{II}$ and $age_\mathrm{III}$ are $\rho=-0.81$,
$\rho = -0.21$ and $\rho = -0.10$ respectively). $\varphi$, the parameter
governing the relative fraction of SSP~II and SSP~III, correlates with
$age_\mathrm{I}$ ($\rho = 0.30$): this positive correlation reflects the fact
that the measured index profiles must be met by diluting a relatively
younger/older starburst (SSP~I) with a correspondingly younger/older outer SSP
(SSP~II + SSP~III). This implies that our indices constrain only the mean age
of the outer population. For this reason, $\varphi$ anti-correlates with
$age_\mathrm{II}$ ($\rho = -0.16$), because for a given value of $age_\mathrm{I}$
and $age_\mathrm{III}$, the required age of the outer SSP must be met either with
a low fraction of young SSP~II stars (larger $\varphi$) or with a high fraction of
relatively older SSP~II stars (lower $\varphi$). Similar reasoning explains the
positive correlation between $\varphi$ and $age_\mathrm{III}$.

In \reffig{f.r.thrage.b} (left panel) we report the posterior distribution
of the three SSPs: it can be seen that SSP~II (dashed yellow histogram) overlaps
with SSP~III (dotted red histogram), but this does not mean that $age_\mathrm{II}
\geq age_\mathrm{III}$, as can be seen from the joint probability distribution
of $age_\mathrm{II}$ and $age_\mathrm{III}$ (third row, second column of the
corner diagram, \reffig{f.r.thrage.a}). It is instead true that SSP~II is
occasionally younger than SSP~I (dashed yellow and solid blue histograms,
respectively): this occurs with a
probability $P<0.07$. However, even these 7\% of cases do not contradict the main
conclusion that PSB galaxies have inverse age gradients. In fact, what really
matters is the mean age of the outer SSP, consisting of both SSP~II and the older
SSP~III. The age of the outer SSP is illustrated in the right panel of
\reffig{f.r.thrage.b} (dot-dashed red histogram), where we reproduce again the
histogram of $age_\mathrm{I}$ for ease of comparison (solid blue line). In no
case we find that the outer age is younger than the $age_\mathrm{I}$ (the
overlap between the two histograms does not take into account the positive
correlation between the age of the three SSPs).
By comparing this histogram with the left panel of \reffig{f.r.psbmodel.b},
illustrating the results for the benchmark model, we see that the three-SSP
model allows for much older outer SSPs than the benchmark model, reflecting the
oversimplification of using a single SSP (in the benchmark model) instead of an
extended star-formation history. Thus in general we can expect older ages for
all of our non-starburst components.
\section{Discussion}\label{s.ds}

The different mass-size relations of SF and Q galaxies, as well as their
evolution with cosmic time, require a link between star formation and structural
evolution in galaxies (\S\ref{s.i}).
There are several physical processes which can cause a SF galaxy to become
quiescent, and each one of them might impart different structural signatures on
newly-quiescent galaxies. For this reason, we can reasonably expect to learn
something about how galaxies become quiescent by studying the structural
differences between SF, Q and newly-quiescent galaxies.
We can roughly divide quenching mechanisms in two classes, based on their
timescale relative to the visibility time of PSB galaxies, strongly constrained
by the lifetime of A-type stars ($<1 \, \mathrm{Gyr}$).

Slow quenching processes act over a few $\mathrm{Gyr}$, longer than the typical
star-formation timescale at $z=0.8$ \citep[defined as the typical inverse specific
star-formation rate $sSFR^{-1} \approx 1 \, \mathrm{Gyr}$;][]{noeske+2007}. 
These processes include: (i) virial shocks \citep[which prevent the accretion of cold
gas, but leave the existing gas disc intact;][]{birnboim+dekel2003,
dekel+birnboim2006}, (ii) radio-mode feedback due to active galactic
nuclei\footnote{but notice that, depending on the angle of the radio jet with respect
to the gas disc, radio-mode feedback may lead to molecular outflows, see e.g.
\citealp{garcia-burillo+2014, sakamoto+2014, morganti+2015, dasyra+2016}}
\citetext{AGN; \citealp{croton+2006}, \citealp{barisic+2017}} and (iii)
stabilization of the gas disc against
fragmentation \citep[$Q-$quenching; ][]{martig+2009, cacciato+2012, forbes+2014,
krumholz+thompson2013, dekel+burkert2014}. These mechanisms cause little or no
disruption to the gas that is currently fueling star formation, so that the
galaxy can continue on the star-forming sequence for some time, until the
cold-gas supply is either exhausted or otherwise unable to form stars. By
definition, these mechanisms act gradually, thereby leaving a Q galaxy with
roughly the same mass and size, and the same structure as the original SF galaxy.
Most SF galaxies form in an \textit{inside-out} fashion \citep{pezzulli+2015,
ellison+2018, wang+2019}, leading to negative age gradients. At the same time,
chemical enrichment models predict negative or flat stellar metallicity gradients
\citep[with some inversion in the centre, see][]{schonrich+mcmillan2017}. Thus the
fact that our control sample of Q galaxies shows negative age and metallicity gradients
(\S\ref{s.2ssp.ss.control}) is qualitatively consistent
with the slow quenching and subsequent passive evolution of SF galaxies from
earlier epochs, without major structural changes. There are two important caveats
to this conclusion. Firstly, our results are derived from stacks, so they are valid
only in an average sense: we cannot say whether (or what fraction of) passive 
galaxies had more complex star-formation histories, with star-formation ending last
in the centre, or with later central starbursts \citep[rejuvenation; see][for an
integrated analysis using LEGA-C]{chauke+2018}. Secondly, and more importantly,
we remark that control galaxies became quiescent at earlier epochs compared to
$z\approx0.8$ PSB galaxies, at a time when the star-formation timescale was shorter.

On the other hand, fast quenching processes happen on relatively short
timescales \citep[$\lesssim 1 \, \mathrm{Gyr}$, e.g.][]{kaviraj+2007,
dekel+burkert2014},
shorter than the timescale of star formation. These
processes involve, in one way or another, the removal of the currently
star-forming gas: through AGN-driven galactic-scale winds
\citep{springel+2005, kaviraj+2007, baron+2018}, through ram-pressure stripping
\citep[in galaxy clusters;][]{gunn+gott1972}\footnote{Given that our sample
consists mostly of central/isolated galaxies, we ignore ram-pressure stripping
from here on.}, or via rapid gas-consumption in
gas-rich mergers \citep{barnes+hernquist1991, barnes+hernquist1996,
hopkins+2009a, dekel+burkert2014}.
A key property of the fast quenching processes is that all of them either
require or cause the presence of centrally-concentrated cold gas. This gas
builds up a dense stellar core, thereby increasing the stellar mass and
shrinking the half-light radius of the underlying galaxy just before it becomes quiescent.
Some of the fast-quenching mechanisms also produce an increasing age trend with
radius \citep{mihos+hernquist1994, bekki+2005}, a signature opposite to the negative
age gradients expected from inside-out formation.\\

\subsection{Evidence of central starbursts in PSB galaxies}\label{s.ds.ss.cburst}

Post-starburst galaxies are passive galaxies with a significant fraction of
young stars. They lack current star formation on timescales of $10\text{-}100 \,
\mathrm{Myr}$ (depending on the tracer used) and they have prominent Balmer
spectral features typical of A-type stars, which have lifetimes of $< 1 \,
\mathrm{Gyr}$. These two timescales constrain the quenching timescale of PSB galaxies,
placing them decisively in the fast-quenching channel. We argue that the
structural properties of PSB galaxies are in qualitative agreement with the
expectations of the fast-quenching scenario described above.

Firstly, our PSB galaxies show a high merger fraction ($0.40$) relative to field
galaxies \citep[$0.16\pm0.02$ at $z\approx1$][their Table 2]{mantha+2018}. Given
our sample size and assuming Poisson uncertainties, the probability that the PSB
sample has the same merger rate as observed in the field is relatively modest,
$P=0.02$. This suggests that gas-rich mergers are an important trigger of PSB
evolution, in agreement with the theoretical scenario proposed above.
Notice, however, that the value $P=0.02$ assumes that the merger classification
was perfect. In practice, there are two main uncertainties that are hard to
quantify in our small sample: misclassification can bias the $P$-value to lower
values (e.g. asymmetries due to dust and chance alignments mis-classfied as
mergers), whereas uncertainties on the visibility time of merger signatures bias
the $P$-value to higher values.
Nevertheless, our results are qualitatively unchanged if we repeat the test
separately for merging and non-merging PSBs (\S\ref{s.r}), therefore it is
unlikely that the high-visibility mergers we are able to discern are the only
channel for PSB formation: the remaining $60\%$ of PSB galaxies that show no
evidence of major mergers could either: i) be mergers below our detection
threshold \citetext{e.g. \citealp{zabludoff+1996}, \citealp{goto2005},
\citealp{yang+2008}, \citealp{pracy+2009} find merger
fractions between 15 and 70 per cent, depending on sample selection and image
quality, see \citealp{pawlik+2018} for a discussion}, or ii) have accreted
low-stellar-mass, gas-rich satellites, or iii) have undergone a qualitatively
different mode of gas accretion, such as cold flows \citep{dekel+2009}.
Regardless, all of these channels are consistent  with the fast quenching scenario.
This high but not overwhelming merger fraction is consistent with the prediction
of cosmological N-body simulations, which suggest roughly half of PSB galaxies
did not experience a recent merger \citep{wilkinson+2018, davis+2019, pawlik+2019}.

Secondly, stacking analysis of $z>1$ PSB galaxies show evidence of high-velocity
outflows \citep{maltby+2019}, capable of removing the star-forming gas and thus
halting star formation abruptly. There is at least one case where this process
has been observed in act \citep{baron+2018}. These outflows may temporarily
delay star formation, until other physical processes, acting on longer
timescales, can permanently halt the supply of cold gas and make the galaxy
quiescent.

Moreover, the structural changes expected from fast quenching go in the right
direction \textit{vis-\`{a}-vis} the observed  differences between normal and
PSB galaxies on the mass-size plane: at a given stellar mass, PSB galaxies are
slightly smaller than normal Q galaxies and are much smaller than
SF galaxies \citep{whitaker+2012, yano+2016, wu+2018}.

Finally, in this work we provide direct evidence that PSB galaxies at
$z\approx0.8$ have younger stellar populations in their centres, thereby
demonstrating that, in these galaxies, star formation occurred last in the centre.
This event is different from the typical inside-out formation that characterises
both the control sample (\reffig{f.r.indrad}, magenta) as well as the bulk of local
SF galaxies \citep[e.g.][]{gonzalezdelgado+2015, zibetti+2017, ellison+2018}.

\subsection{The progenitors of PSB galaxies}\label{s.ds.ss.progs}
The discovery that PSB galaxies have inverse stellar population gradients
is in agreement with the hypothesis that PSB galaxies are descendants of
compact star-forming (CSF) galaxies.
These galaxies are characterised by high star-formation rates and small size
relative to SF galaxies of the same mass \citep[star-formation rate $> 100 \,
\mathrm{M_\odot \, yr^{-1}}$, $R_e < 3 \, \mathrm{kpc}$; e.g.][]{barro+2013,
barro+2014, vandokkum+2015}. Yet this high star-formation rate is mostly
undetected in the rest-frame optical wavelengths because it is obscured by
large optical depth.

Even though the number density of CSF galaxies is too high compared to
that of PSB galaxies \citep{wild+2016}, CSF galaxies are not guaranteed
to become quiescent, hence their number density does not need to match exactly
that of PSB galaxies. For example, CSF galaxies could transition to a state of
lowered star-formation rate instead of becoming quiescent.

On the other hand, there is evidence that CSF galaxies, too, have inverse
stellar population gradients, because their star formation is less
extended than the already-formed stellar population \citep{barro+2016,
barro+2017, tadaki+2017, popping+2017}. If CSF galaxies transitioned to
a relatively long phase of lowered star-formation rate, characterised by
inside-out growth, the inverse signature on their stellar populations might
be erased \citep[at least in part, giving rise to U-shaped age profiles observed
in some local early-type galaxies][]{zibetti+2020}.
If at least some of these galaxies were about to undergo a rapid cessation of
star formation, the inverse gradients would be ``frozen''. Our finding that
PSB galaxies harbour inverse stellar population gradients agrees with the
hypothesis that at least some of their progenitors might be CSF galaxies that
underwent rapid quenching.\\

Alternatively, the progenitors of PSB galaxies could be galaxies detected
at sub-mm wavelength \citep[sub-mm galaxies][]{smail+1997, hughes+1998,
barger+1998, eales+1999}. These galaxies are characterised by high
star-formation rates \citep{swinbank+2014}, consistent with the star-formation
history of PSB galaxies \citep{wild+2020}. Moreover, the combination of the
number density of sub-mm galaxies \citep[$\approx 6$ times higher than that of
PSB][]{swinbank+2014, simpson+2014} and the visibility time of the sub-mm phase
\citep[$\approx 6$ times shorter than that of PSB][]{hainline+2011, hickox+2012}
cancel out and match the observed number density of observed PSB galaxies
\citep{wild+2020}.

\subsection{\textit{Caveats}}

Given our sample selection criteria, it is difficult to assess the generality
of our results \textit{vis-\'a-vis} the unbiased population of PSB galaxies at
redshift $z\approx0.8$. In fact, our SNR and spatial resolution bias the sample
towards the largest and brightest PSB galaxies, therefore a trend between
stellar-population gradients and size or luminosity could mean that our
conclusions are not representative of the whole PSB population. Conversely,
removing galaxies with small \re means that our sample is biased against the
most compact PSB galaxies, which may represent the youngest galaxies in this
category \citep{wu+2020}. For these reasons, we cannot draw
stringent conclusions concerning how size affects the properties of our
galaxies (but we remark that jackknifing our sample about the median \re
yields consistent results for the smallest and largest PSB considered here).
However, we can still robustly conclude that inverse age gradients are
representative of $\approx 50\%$ of the PSB population \citep[even
though individual PSB galaxies can indeed have inside-out gradients, e.g. 258467
has a positive \hdeltaa gradient with four standard deviations; see also
Appendix~\ref{app.colour} and][]{hunt+2018}.
Moreover, our data suggest that, provided the galaxies are at least partially
resolved, there is no strong dependence of the incidence of inverse gradients
with either galaxy size or galaxy concentration.
Finally, we find that our results cannot be explained by environment effects
(i.e. ram-pressure stripping), because the inverse gradients persist for
central/isolated galaxies only.

Using the UVJ diagram to select Q galaxies may bias our sample, by
including misclassified SF galaxies. There are two indications that
this is not the case. Firstly, if we select only PSB galaxies where the
wavelength range includes rest-frame $\mathrm{H\beta}$, and if for these
galaxies we further require equivalent width $EW(\mathrm{H\beta}) \geq 1 \,
\text{\AA}$, we still find inverse gradients for \hdeltaa (emission lines
have negative EW). For \fe4383 the gradients are consistent with being flat, but this
is likely due to the smaller sample size. Secondly, a preliminary analysis of SF
galaxies finds inside-out gradients, similar to non-PSB quiescent galaxies,
therefore any contamination would make our results weaker.

Another caveat is that, at least for PSB galaxies, the SNR of \hdeltaa is
higher than the SNR of \fe4383, which might skew our analysis. Furthermore,
we are unable to constrain the abundance of $\alpha$-elements relative to
$\mathrm{Fe}$, which is important to properly constrain the metallicity of
a stellar population \citep[e.g.][]{gonzalez1993, thomas+2003, conroy2013},
We note though that this should not be a concern for determining the age of the
PSB galaxies \citep{leonardi+rose1996}, but might affect the older control
sample.

Unfortunately at present we cannot derive strong quantitative constraints on the
stellar population gradients in PSB galaxies: seeing convolution prevents us
from breaking the degeneracy between spatial distribution, age and mass fraction
of the young, central sub-population. Still, the observed agreement with local
observations is promising (\refsec{s.2ssp.ss.benchmark}).
In addition, the need for stacking data from different galaxies means that
our models do not necessarily capture the unbiased average of the population.
Individual galaxies show a plethora of radial behaviours, as suggested by
resolved studies of individual galaxies in the local Universe \citep[e.g.][]{
pracy+2013, owers+2019, chen+2019}: the fact that some PSB galaxies have flat
gradients is not in contradiction with our results. These different radial
behaviours may depend on the timing and on the properties of the burst.

\subsection{Future work}

Current observing facilities are unable to disentangle this degeneracy,
therefore we foresee a multi-pronged approach.

Space-based observations have enough spatial resolution to resolve the inner
structure of PSB galaxies. Unfortunately, at the time of this writing, the
number of PSB galaxies with adequate photometry in two bands is limited,
and cannot be used to draw strong conclusions for our sample
(Appendix~\ref{app.colour}). At variance with our findings, \citet{maltby+2018}
find no evidence of colour gradients in a sample of 80 PSB galaxies at $0.5 < z
< 2$, but they rely on photometry only, therefore a direct comparison is not
warranted.

Another implication of the scenario proposed is that the youngest PSB galaxies
ought to have the smallest half-light radii. This is because the younger
central population has a lower mass-to-light ratio, but dims faster than the
older, extended population \citetext{as indeed confirmed by \citealp{wu+2020};
see their discussion for the effect of dust}.

Studying the structure of local galaxies is complicated by the intervening
evolution between $z\approx0.8$ and now, but inverse age structures have been
found in the most massive early-type galaxies \citep{zibetti+2020}. This
approach should be combined with a study of star-burst dwarf galaxies in the
local Universe. These systems have many properties in common with higher-mass
galaxies in the early Universe \citep[e.g.][]{lelli+2014, yang+2017}.
Intriguingly, some of these local galaxies have inverse age structures caught
in formation \citep{zhang+2012}, and might shed light on the physical processes
occurring at high gas fractions.\\

Accurate modelling involving more than indices \citep[e.g.][]{zibetti+2017} may
also provide more stringent constraints on the age and metallicity of PSB
galaxies, resolving some of the degeneracies inherent to using only two indices
(\S\ref{s.2ssp.ss.benchmark}).
The upcoming third data release of LEGA-C will likely improve both the precision
and the accuracy of our measurements. Moreover, owing to a larger sample and
thanks to the improved sky subtraction at the reddest wavelengths, it will unlock
additional stellar-population information.

\section{Summary}

We have used a sample of seventeen UVJ-colour selected, spectrally-confirmed
post-starburst (PSB) galaxies from the deep, $z \approx 0.8$ LEGA-C Survey, to
study the radial variation of the strength of the \hdeltaa and \fe4383
absorption indices. We find that:
\begin{itemize}
    \item PSB galaxies show negative \hdeltaa radial gradients,
    and positive \fe4383 gradients ($P$-values $10^{-4}$ and $0.04$
    respectively).
    \item the control sample of quiescent, non-PSB galaxies, selected to match
    the mass range of the PSB sample, presents positive \hdeltaa radial gradients
    and negative \fe4383 radial gradients ($P$-values $0.07$ and $0.02$
    respectively)
    \item these trends imply a positive (inverse) age gradient with radius
    for PSB galaxies, and a negative age and/or metallicity
    gradient for control quiescent galaxies (the control sample).
    \item coupled with the mass and size distribution of PSB and quiescent
    galaxies, and with the higher-than-average merger fraction ($0.4$), our data
    suggest that PSB galaxies have undergone a central starburst, which
    decreased their effective radius and possibly caused them to become quiescent
    on a fast timescale ($<1 \, \mathrm{Gyr}$).
\end{itemize}

\section*{Acknowledgments}

We thank the anonymous referee for helpful comments and suggestions. Based on
observations made with ESO Telescopes at the La Silla Paranal
Observatory under programme ID 194-A.2005 (The LEGA-C Public Spectroscopy
Survey). This project has received funding from the European Research Council
(ERC) under the European Union’s Horizon 2020 research and innovation programme
(grant agreement No. 683184). 
We are grateful to N.~Scott for sharing the spectral fitting code.
PFW acknowledges the support of the East Asian Core Observatories Association fellowship.
TMB is supported by an Australian Government Research Training Program Scholarship.
AG and SZ acknowledge support from Istituto Nazionale di Astrofisica (PRIN-SKA2017
program 1.05.01.88.04)

This work made extensive use of the Debian GNU/Linux operative system, freely
available at http://www.debian.org. We used the Python programming language
\citep{vanrossum1995}, maintained and distributed by the Python Software
Foundation, and freely available at http://www.python.org. We acknowledge the
use of {\sc scipy} \citep{jones+2001}, {\sc matplotlib} \citep{hunter2007}, {\sc emcee}
\citep{foreman-mackey+2013}, {\sc astropy} \citep{astropyco+2013} and {\sc pathos}
\citep{mckerns+2011}.
During the preliminary analysis we have made extensive use of {\sc TOPCAT}
\citep{taylor2005}.

\section*{Data availability}

The raw data used in this work is available in the public domain, through the
\href{http://archive.eso.org}{ESO Science Archive} and through the \href{
https://archive.stsci.edu/hst/}{Mikulski Archive for Space Telescopes (MAST)}.
Part of the reduced data has been released with the second public data release
of LEGA-C \citep{straatman+2018}, available on the
\href{https://www2.mpia-hd.mpg.de/home/legac/index.html}{LEGA-C website}.
Resolved measurements will be available in the upcoming third public data
release of LEGA-C.



\bibliographystyle{mnras}
\bibliography{astrobib}



\appendix

\section{\hgammaa and \dn4000 indices}\label{app.otherinds}

\begin{figure}
  \includegraphics[type=pdf,ext=.pdf,read=.pdf,width=1.0\columnwidth]{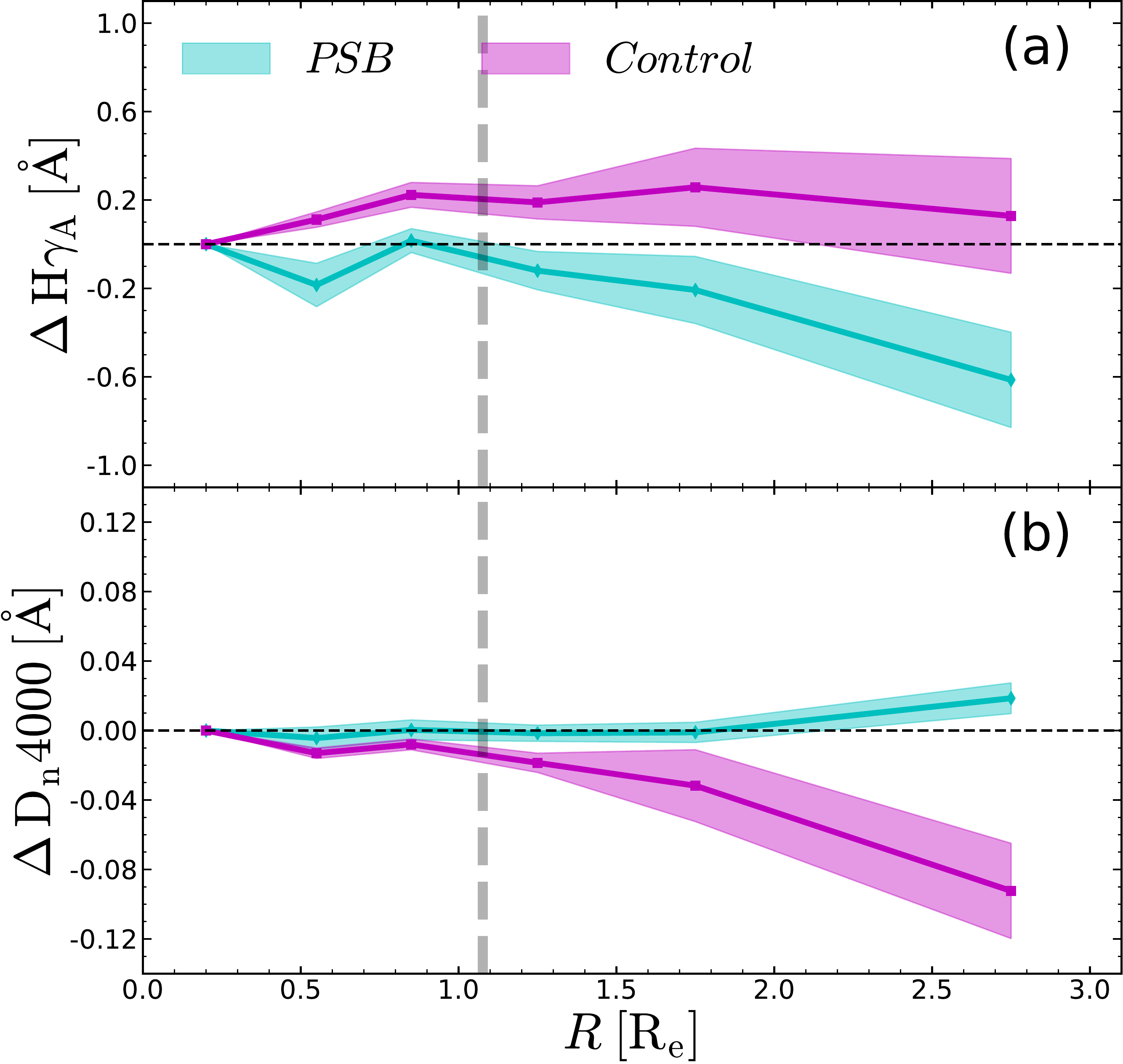}
  {\phantomsubcaption\label{f.app.indrad.a}
   \phantomsubcaption\label{f.app.indrad.b}}
  \caption{Same as \reffig{f.r.indrad}, but for \hgammaa
  (panel~\subref{f.app.indrad.a}), and \dn4000 (panel~\subref{f.app.indrad.b}).
  These two indices show again that PSB galaxies (cyan) have different radial
  profiles than the control sample of quiescent, non-PSB galaxies (magenta). For
  PSB galaxies, \hgammaa is radially decreasing and \dn4000 is flat, whereas for
  control galaxies, \hgammaa is radially increasing and \dn4000 is radially
  decreasing.
  }\label{f.app.indrad}
\end{figure}

The PSB and control sample have qualitatively different radial trends
also in their \dn4000 and \hgammaa indices (\reffig{f.app.indrad})

\section{Colour gradients}\label{app.colour}

Depending on the strength of the intrinsic age gradient, the presence of
inverse stellar gradients in PSB galaxies can be investigated using
space-based photometry. For a meaningful test, we require two conditions:
firstly, we need space-based photometry to derive intrinsic radial profiles of
rest-frame optical colour. Secondly, we need to compare these radial colour
profiles to the index profiles of the relevant galaxies. A meaningful test also
requires that the index profiles are at least tentatively detected: profiles
that are consistent with being flat contain no information. In particular, we
seek to match positive colour gradients (outskirts redder than the centre) to
negative \hdeltaa gradients (outskirts older/more metal-rich than the centre). 
For simplicity, we refer to both positive colour gradients and negative \hdeltaa
gradients as inverse gradients, and vice versa, negative colour gradients and
positive \hdeltaa gradients are inside-out gradients.

\begin{figure}
  \includegraphics[type=pdf,ext=.pdf,read=.pdf,width=1.0\columnwidth]{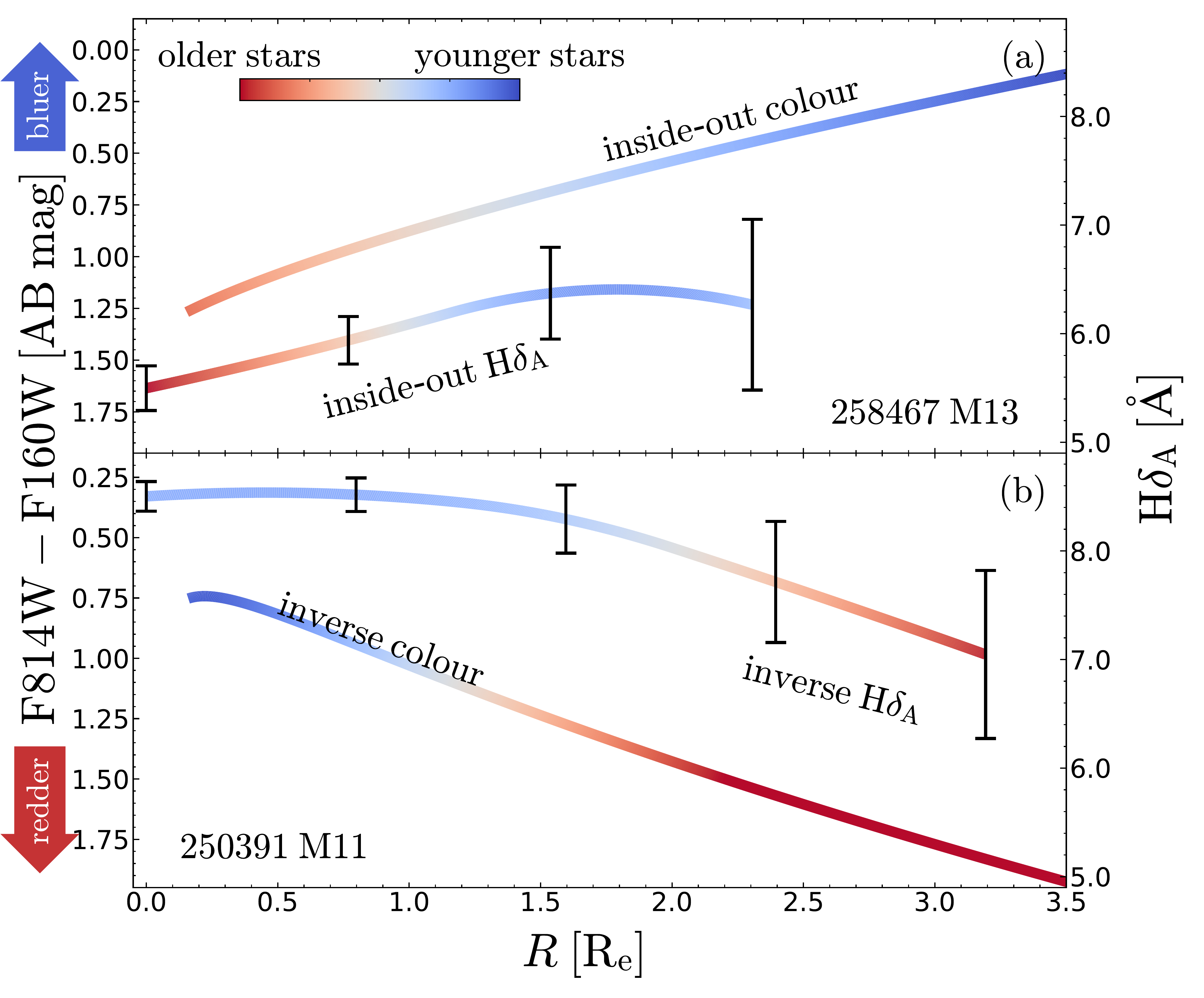}
  {\phantomsubcaption\label{f.app.colour.a}
   \phantomsubcaption\label{f.app.colour.b}}
  \caption{We show the clear connection between rest-frame optical colour
  profiles (lines with no error bars, left-hand scale) and \hdeltaa profiles
  (lines with error bars, right-hand scale), for the two PSB
  galaxies with both available colour profiles \textit{and} significant
  \hdeltaa gradients. 258467 M13 shows clear inside-out structure (outskirts bluer
  and with stronger \hdeltaa than the centre; Panel~\subref{f.app.colour.a}),
  whereas 250391 M11 shows clear inverse structure (outskirts redder and with
  weaker \hdeltaa than the centre; Panel~\subref{f.app.colour.b}). The lines
  are arbitrarily colour-coded so that bluer optical colour and stronger
  \hdeltaa absorption, corresponding to relatively younger stars, are rendered
  by bluer colour hues.}\label{f.app.colour}
\end{figure}

Unfortunately, at the time of this writing, we can only present data for two
galaxies (\reffig{f.app.colour}). The proposed approach is in fact limited by
the unavailability of suitable data. As a precondition, we have reliable COSMOS
F814W photometry for all the PSB galaxies in our sample, but, in general,
suitable redder-wavelength images are unavailable. When available, these images
have broader PSF than F814W photometry, and as a result, raw images show an
artificially high fraction of galaxies with blue centers. Therefore we require
also a reliable PSF deconvolution, which further limits the sample size.
The overlap with the CANDELS Survey is insufficient \citep{grogin+2011}: out of
seventeen PSB galaxies considered here, only galaxy 211263~M1 is present in
CANDELS. This galaxy (not pictured) has flat \hdeltaa profile and negative colour
gradient, and is therefore uninformative.

The COSMOS-DASH Survey \citep[hereafter: DASH;][]{mowla+2019}, provides HST
F160W imaging of the COSMOS field. Given that LEGA-C is selected from the COSMOS
field, and that DASH currently covers $\approx30\%$ of this field, DASH data are
available for five of our PSB galaxies. Of these, one is galaxy 211263~M1, which
we have already discarded. Of the remaining four galaxies, two more have flat
\hdeltaa gradient (110805 M3 and 216730 M9); moreover, both of these galaxies
have unreliable {\sc galfit} fits, in that the best-fit parameters reached
the limits of the allowed range. The reason is probably that DASH is less deep
than either COSMOS or CANDELS: the five $\sigma$ point-source detection limit
for DASH is $H = 25.2 \, \mathrm{mag}$ \citep{mowla+2019}, whereas CANDELS has
$H = 27 \, \mathrm{mag}$ \citep{grogin+2011} and COSMOS has $I = 27.2 \,
\mathrm{mag}$ \citep{koekemoer+2007}.
We are left with only two PSB galaxies. The first (258467 M13;
Panel~\subref{f.app.colour.a}) has both inside-out colour and inside-out
\hdeltaa profile, whereas the second (250391 M11; Panel~\subref{f.app.colour.b}),
has both inverse colour and inverse \hdeltaa profile.

The agreement between the two reliable index gradients with the
respective colour gradients is a reassuring consistency check for our
methodology. We conclude that our detection of an inverse age
gradient in the typical PSB is a robust result, and that space-based photometry
may represent an alternative mean to investigate the inverse or inside-out
structure of PSB galaxies.

\bsp	
\label{lastpage}
\end{document}